\documentclass[12pt]{article}
\usepackage{amsmath}
\usepackage{amssymb,amscd}
\usepackage{bm}
\usepackage{wrapfig}

\usepackage{color}
\usepackage{bm}
\usepackage[all]{xy}

\usepackage{theorem}

\newtheorem{theorem}{Theorem}[section]
 \newtheorem{proposition}[theorem]{Proposition}
 \newtheorem{corollary}[theorem]{Corollary}

{\theorembodyfont{\normalfont}
 \newtheorem{definition}[theorem]{Definition}
 \newtheorem{example}{Example}[section]
 
 }

\newcommand{\qed}{\nobreak \ifvmode \relax \else
      \ifdim\lastskip<1.5em \hskip-\lastskip
      \hskip1.5em plus0em minus0.5em \fi \nobreak
      \vrule height0.75em width0.5em depth0.25em\fi}


\newlength{\minitwocolumn}
\usepackage{amsmath}
\usepackage{amssymb,amscd}
\usepackage{bm}
\usepackage{amsbsy} 
\usepackage{wrapfig}
\usepackage{theorem}

\newcommand{\beq}{\begin{equation}}
\newcommand{\eeq}{\end{equation}}
\newcommand{\bea}{\begin{eqnarray*}}
\newcommand{\eea}{\end{eqnarray*}}
\newcommand{\beqa}{\begin{eqnarray}}
\newcommand{\eeqa}{\end{eqnarray}}

\def\bR{{\mathbb{R}}}

\def\bZ{{\mathbb{Z}}}

\newcommand{\calG}{{\mathcal G}}

\newcommand{\calL}{{\mathcal L}}
\newcommand{\calM}{{\mathcal M}}

\newcommand{\Map}{{\rm Map}}

\newcommand{\sbv}[2]{{\{{{#1},{#2}}\}}}

\newcommand{\courant}[2]{{[{{#1},{#2}}]_D}}

\newcommand{\bracket}[2]{\langle #1,\,#2\rangle}

\newcommand{\omegac}{\omega_{can}}
\newcommand{\rd}{\mathrm{d}}
\def\sd{{\bm{\mathrm{d}}}}
\def\bbd{{\bm{\mathrm{d}}}}

\newcommand{\Fsx}{F_{x}{}}
\newcommand{\Fsxi}{F_{p}{}}
\newcommand{\Fsq}{F_{\eta}{}}

\newcommand{\Fse}{F_{e}{}}
\newcommand{\Fsqe}{F_{Qe}{}}

\newcommand{\htp}{h{}}

\def\rc#1{{\color{red}#1}}

\begin{document}

%%%%%%%%%%%%%%%%%%%%%%%%%%%%%%%%%%%%%%%%%%%%%%%%%%%%%%%%%%%%%%%%%%
%%%%%%%%%%%%%%%%%%%%%%%% Title %%%%%%%%%%%%%%%%%%%%%%%%%%%%%%%%%%%
%%%%%%%%%%%%%%%%%%%%%%%%%%%%%%%%%%%%%%%%%%%%%%%%%%%%%%%%%%%%%%%%%%

\baselineskip 0.7cm

\begin{titlepage}
%\today
\begin{flushright}
%MISC-2011-
\end{flushright}

\vskip 1.35cm
\begin{center}
{\Large \bf
Momentum section on Courant algebroid 
and constrained Hamiltonian mechanics
}
\vskip 1.2cm
Noriaki Ikeda
%${}^b$
\footnote{E-mail:\
nikedaATse.ritsumei.ac.jp
%,
%ikeda@yukawa.kyoto-u.ac.jp
}
\vskip 0.4cm

{\it
%${}^b$
Department of Mathematical Sciences,
Ritsumeikan University \\
Kusatsu, Shiga 525-8577, Japan \\
}
\vskip 0.4cm

\today

\vskip 1.5cm

\begin{abstract}
We propose a generalization of the momentum map on a symplectic manifold with 
a Lie algebra action to a Courant algebroid structure. 
The theory of a momentum section on a Lie algebroid is generalized to the theory compatible with a Courant algebroid.
%A Hamiltonian Lie algebroid is generalized to a Hamiltonian Courant algebroid.
As an example, we identify the momentum section in a constrained Hamiltonian mechanics with Courant algebroid symmetry.
Moreover, we construct cohomological formulations by considering the BFV and BV formalism of this Hamiltonian system.
The Weil algebra for this structure is constructed.
%Equations of inhomogeneous parts of the BFV functional and the Hamiltonian, or the BV functional are identified to a momentum section.
\end{abstract}
\end{center}
\end{titlepage}

\tableofcontents

\setcounter{page}{2}

%%%%%%%%%%%%%%%%%%%%%%%%%%%%%%%%%%%%%%%%%%%%%%%%%%%%%%%%%%%%%%%%%%
%%%%%%%%%%%%%%%%%%%%%%%% Article %%%%%%%%%%%%%%%%%%%%%%%%%%%%%%%%%
%%%%%%%%%%%%%%%%%%%%%%%%%%%%%%%%%%%%%%%%%%%%%%%%%%%%%%%%%%%%%%%%%%

\rm

%%%%%%%%%%%%%%%%%%%%%%%%%%%%%%%%%%%%%%%%%%%%%%%%%%%%%%%%%%%%%%%%%%%%%
%%%%%%%%%%%%%%%%%%%%%%%%%%%%%%   SEC      %%%%%%%%%%%%%%%%%%%%%%%%%%
%%%%%%%%%%%%%%%%%%%%%%%%%%%%%%%%%%%%%%%%%%%%%%%%%%%%%%%%%%%%%%%%%%%%%
\section{Introduction}
\noindent
Interesting structures that generalize a Lie group and a Lie algebra in Poisson geometry are a Lie algebroid and a Lie groupoid. (for example, see a textbook \cite{Mackenzie}) 
As a further generalization, 
%a Courant algbroid \cite{LXW} appears in many contexts as a fundamental structure.
a Courant algebroid \cite{LWX} which is a 2-categorical generalization of a Lie algebroid appears in the analysis of the Dirac structure \cite{Courant}, and a generalization of a Lie bialgebra and Poisson Lie group theory, generalized geometry \cite{Hitchin:2004ut, Gualtieri:2003dx}, a topological sigma model \cite{Ikeda:2002wh, Roytenberg:2006qz}, T-duality in string theory \cite{Cavalcanti:2011wu}, etc.

A geometric structure was discovered as compatibility conditions of gauge theories and Hamiltonian mechanics with a Lie algebroid structure \cite{Mayer:2009wf, Kotov:2014iha, Kotov:2016lpx, Ikeda:2018rwe}.
A moment(um) map theory is a fundamental theory in symplectic geometry
inspired by the mechanics with a Lie group action.
Recently, Blohmann and Weinstein \cite{Blohmann:2018} have proposed the equivalent structure as a generalization of a momentum map and a Hamiltonian $G$-space with a Lie algebra action to the Lie algebroid setting, inspired by analysis of the Hamiltonian formalism of the general relativity \cite{Blohmann:2010jd}. 
They are called a momentum section and a Hamiltonian Lie algebroid.
The author gave new examples in the constrained Hamiltonian mechanics 
and sigma models, and generalized a momentum section theory to a pre-multisymplectic manifold \cite{Ikeda:2019pef}. 
%The equivalent structure has been proposed as the moment(um) map theory 
%
There are related papers \cite{Glowacki, Blohmann:2021}.

In this paper, we generalize a momentum section theory on a pre-symplectic manifold compatible with a Lie algebroid to a Courant algebroid. We obtain conditions similar to a Lie algebroid, however geometric quantities of the Lie algebroid are replaced to ones of the Courant algebroid.
We define a Hamiltonian Courant algebroid as a generalization of the Hamiltonian $G$-space.

We show that a simple constrained Hamiltonian mechanics system is an example of our proposal.
The compatibility condition of the constrained Hamiltonian mechanics on a cotangent bundle $T^*M$ with a Courant algebroid $E$ over a base manifold $M$ gives
existence of a momentum section.
%In a momentum section on a Lie algebroid, we discussed similar mechanics
%in the papers \cite{Ikeda:2018rwe, Ikeda:2019pef}.
Essentially, the 0-th order term with respect to the momentum $p$ in the constraint is identified to a momentum section. 
The conditions are realized as Poisson brackets of constraints and the Hamiltonian.
This example is a very natural physical system, thus, a momentum section can be understood as a natural geometric structure in physical theories.

In the preceding sections, we consider cohomological realizations of a momentum section and a Hamiltonian Courant algebroid, which give clear interpretations of complicated conditions.
We consider the BFV formalism and the BV formalism of the above Hamiltonian mechanics.
The BRST-BFV charge functional $S_{BFV}$ in the BFV formalism, and
the BV action functional $S_{BV}$ in the BV formalism are the Hamiltonian for homological vector fields $Q$ such that $Q^2=0$, which are differentials of complexes.
In the BFV formalism, Poisson brackets of two fundamental functions, the BFV charge function $S_{BFV}$ and the BFV Hamiltonian $H_{BFV}$, are equivalent to the consistency conditions of a Hamiltonian Courant algebroid.
In the BV formalism, the BV bracket (an odd Poisson bracket) of the BV action functional $S_{BV}$ is equivalent to the consistency conditions.
As applications, we construct the Weil model and the Cartan model of a Hamiltonian Courant algebroid to apply the equivariant cohomology theory.

This paper is organized as follows.
In Section 2, we give definitions of a momentum section on a Courant algebroid and consider some examples.
In Section 3, We show that a momentum section appears in
a constrained Hamiltonian mechanics with a Courant algebroid structure.
In Section 4, we consider the BFV formalism of the Hamiltonian mechanics
in Section 3.
In Section 5, the BV formalism is constructed based on the FHGD formulation
In Section 6, the Weil algebra and the Cartan model are discussed.
Section 7 is devoted to discussion and outlook.
In Appendix, some formulas are summarized.

\textit{Note}: After this article has been completed, 
we were informed that Hancharuk and Strobl considered a Lie 2-algebroid 
extension of the same constrained mechanics, which is different but complementary generalization from our results. 
See \cite{Hancharuk-Strobl}.

%%%%%%%%%%%%%%%%%%%%%%%%%%%%%%%%%%%%%%%%%%%%%%%%%%%%%%%%%%%%%%%%%%%%
%%%%%%%%%%%%%%%%%%%%%%%%%%%%%%%%%%%%%%%%%%%%%%%%%%%%%%%%%%%%%%%%%%%%
%%%%%%%%%%%%%%%%%%%%%%%%%%%%%%%%%%%%%%%%%%%%%%%%%%%%%%%%%%%%%%%%%%%%
\section{Courant algebroid and momentum section}
\subsection{Courant algebroid and $Q$-manifold}
We summarize basic properties of a Courant algebroid used in this paper.
\begin{definition}[Courant algebroid]\cite{LWX}
\label{def:Courant_algebroid}
Let $E$ be a vector bundle over a manifold $M$.
A \textit{Courant algebroid} is a quadruple $(E, [-,-]_D, \rho, \bracket{-}{-})$ where $[-,-]_D$ is a bilinear bracket on $\Gamma (E)$ called the Dorfman bracket, $\rho : E \to TM$ is a bundle map called the anchor map, 
and an inner product $\bracket{-}{-}$ is a non-degenerate bilinear form on $\Gamma (E)$.
They satisfy the following axioms for any $e_i \in \Gamma (E)$ and $f \in C^{\infty}(M)$:
\begin{enumerate}
\item The bracket $[-,-]_D$ satisfies the Leibniz identity 
%\eqref{eq:Leibniz_identity}.
 \begin{align}
 [e_1, [e_2,e_3]_D]_D = [[e_1,e_2]_D,e_3]_D + [e_2,[e_1,e_3]_D]_D.
\label{Leibnizidentity}
 \end{align}
\item $\rho ([e_1,e_2]_D) = [\rho(e_1) , \rho(e_2)]$.
\item $[e_1, f e_2]_D = f [e_1,e_2]_D + (\rho (e_1) \cdot f) e_2 $.
\item $[e,e]_D = \frac{1}{2} \mathcal{D} (e,e)$.
\item $\rho (e_1) \cdot \bracket{e_2}{e_3} = \bracket{[e_1,e_2]_D}{e_3} 
+ \bracket{e_2}{[e_1,e_3]_D}$.
\end{enumerate}
Here $\mathcal{D}$ is a generalized exterior derivative on $\Gamma (E)$
defined by $\bracket{\mathcal{D} f}{x} = \frac{1}{2} \rho(x) f$.
\end{definition}

\begin{example}\label{starndardCA}
Let $M$ be a smooth manifold and take the direct product bundle 
$TM \oplus T^*M$.
For $X + \xi, Y + \eta \in \Gamma(TM \oplus T^*M)$,
we define the inner product,
\begin{align}
\bracket{X + \xi}{Y + \eta} = \iota_X \eta + \iota_Y \xi.
\end{align}
the anchor map
\begin{align}
\rho(X + \xi) f = Xf,
\end{align}
for $f \in C^{\infty}(M)$, 
and the Dorfman bracket,
\begin{align}
[X+\xi, Y + \eta ]_D = [X, Y] + \calL_{X} \eta - d \iota_Y \xi + 
\iota_X \iota_Y h,
\end{align}
for a closed 3-form $h \in \Omega^3(M)$.
Then, $(TM \oplus T^*M, \bracket{-}{-}, \rho, [-,-]_D)$ is a Courant algebroid.
This Courant algebroid is called the \textit{standard Courant algebroid} with $h$-flux.
\end{example}

In order to describe a Courant algebroid as a Q-manifold
\cite{Schwarz:1992nx}, we consider a shifted vector bundle, $E[1]$, which is a graded manifold with a coordinate of the fiber shifted by $1$.
A section on $E^*$ is identified as a function on a graded manifold $E[1]$. 
One can refer to some references of mathematics of 
a graded manifold, Q-manifold related to a Courant algebroid
\cite{Roytenberg99, Cattaneo:2010re, Ikeda:2012pv}

Even and odd local coordinates $(x^i, \eta^a)$ on $E[1]$ are introduced, where $x^i$ is a coordinate on $M$ and $\eta^a$ is a basis of the fiber of degree $1$.
We denote a fiber metric on $E^*$ by $k^{ab} = \bracket{\eta^a}{\eta^b}$.
A Q-manifold structure for a Courant algebroid is defined on 
a graded cotangent bundle $T^*[2]E[1]$ \cite{Roytenberg99}.
Canonical conjugates of $x^i$ and $\eta^a$
are denoted by $p_i$, and $k_{ab} \eta^b$, where we identify $E$ and $E^*$
by the inner product $\bracket{-}{-}$.
We have coordinates $(x^i, p_i, \eta^a)$ of degree $(0, 2, 1)$.

The homological vector field $Q$ for a Courant algebroid on $T^*[2]E[1]$ is
given by
\begin{align}
Q &= \rho^i_a(x) \eta^a \frac{\partial}{\partial x^i}
+ \left(\rho^i_a(x) p_i + \frac{1}{2} f_{acd} \eta^c \eta^d \right) 
k^{ab} \frac{\partial}{\partial \eta^b}
+ \left(\partial_i \rho^j_a(x) p_j \eta^a 
+ \frac{1}{3!} \partial_i f_{abc} \eta^a \eta^b \eta^c \right) 
\frac{\partial}{\partial p_i},
\label{homologicalvf}
\end{align}
where $\rho(e_a) = \rho^i_a(x) \partial_i$ for a basis $e_a$ of $E$, and $f_{abc}$ is a structure function of the Dorfman bracket satisfying $[e_a, e_b]_D = f_{abc} k^{cd} e_d$.
$Q$ is a vector field of degree $1$ such that $Q^2=0$ if and only if $E$ is a Courant algebroid, which define the Courant algebroid differential ${}^E\rd$ on the complex in the space $C^{\infty}(T^*[2]E[1])$.

%%%%%%%%%%%%%%%%%%%%%%%%%%%%%%%%%%%%%%%%%%%%%%%%%%%%%%%%%%%%%%%%%%%%
%%%%%%%%%%%%%%%%%%%%%%%%%%%%%%%%%%%%%%%%%%%%%%%%%%%%%%%%%%%%%%%%%%%%
%%%%%%%%%%%%%%%%%%%%%%%%%%%%%%%%%%%%%%%%%%%%%%%%%%%%%%%%%%%%%%%%%%%%
\subsection{Momentum section and Hamiltonian Courant algebroid}
In this section, we propose
a momentum section of the Courant algebroid and a Hamiltonian Courant algebroid as a generalization of the paper \cite{Blohmann:2018}.
%See also \cite{Ikeda:2019pef}.

%In this section, we summarize the property of $\mu \in \Gamma(E^*)$ obtained from the consistency condition of the constrained Hamiltonian system.

Let $M$ be a pre-symplectic manifold with a pre-symplectic form $B \in \Omega^2(M)$, i.e., a closed $2$-form which is not necessarily nondegenerate. 
We consider a Courant algebroid $(E, [-,-]_D, \rho, \bracket{-}{-})$ over a pre-symplectic manifold $(M, B)$.

We introduce a connection (a linear connection) on $E$. i.e.,
a covariant derivative $D: \Gamma(E) \rightarrow \Gamma(E \otimes T^*M)$, satisfying $D(fe) =f De +  \rd f \otimes e$ for a section $e \in \Gamma(E)$
and a function $f \in C^{\infty}(M)$.
The connection is extended to $\Gamma(M, \wedge^{\bullet} T^*M \otimes E)$
as a degree $1$ operator.
%Let $D$ be the induced exterior covariant derivative.

An $E^*$-valued 1-form $\gamma \in \Omega^1(M, E^*)$ is defined by
\begin{eqnarray}
\bracket{\gamma(v)}{e} = - B(v, \rho(e)),
%\omega_{MS}(v, \rho(a)).
\label{HH0}
\end{eqnarray}
for all sections $e \in \Gamma(E)$ and vector fields $v \in \mathfrak{X}(M)$.
%where $\bracket{-}{-}$ is the inner product of the Courant algebroid.
%Here 
%$B = - \omega_{MS}$, thus, in local coordinates,
%$\gamma_{ia} = - B_{ij} \rho^j_a$
The following two conditions are introduced for $E$ on 
a pre-symplectic manifold $(M, B)$.
\footnote{(H2) and (H3) correspond to the number in \cite{Blohmann:2018}.
(H1) is discussed later.}
\medskip\\
\noindent
(H2)
A section $\mu \in \Gamma(E^*)$ is a \emph{$D$-momentum section} if
it satisfies
\begin{eqnarray}
D \mu =\gamma,
% \gamma,
\label{HH2}
\end{eqnarray}
%
%\medskip\\
%\noindent
(H3)
A $D$-momentum section $\mu$ is \emph{bracket-compatible} if it satisfies
\begin{eqnarray}
&& {}^E\rd \mu (e_1, e_2) = - \bracket{\gamma(\rho(e_1))}{e_2},
\label{HH3}
\end{eqnarray}
for all sections $e_1, e_2 \in \Gamma(E)$.
A $D$-momentum section on a Lie algebroid has formally the same equation as Equation \eqref{HH3}
however the $E$-differential ${}^E d$ is different.
In Equation \eqref{HH3}, ${}^E d$ is the Courant algebroid differential induced from $Q$ in \eqref{homologicalvf}.
Especially, note that
\begin{eqnarray}
&& {}^E\rd \mu (f) = \rho(\mu^*)f,
\label{HH4}
\end{eqnarray}
for $f \in C^{\infty}(M)$ is automatically satisfied for the Courant algebroid differential ${}^E \rd$.
Here the dual momentum section $\mu^* \in \Gamma(E)$ is defined by $\mu^*(e) = \bracket{\mu}{e}$.

A Hamiltonian Courant algebroid is defined as follows.
\begin{definition}\label{weaklyHamiltonianCA}
A Courant algebroid $E$ with a connection $D$ and a section $\mu \in \Gamma(E^*)$
is called \textbf{weakly Hamiltonian} if (H2) is satisfied.
%If the condition is satisfied on a neighborhood of every point in $M$, 
%it is called locally weakly Hamiltonian.
\end{definition}

\begin{definition}\label{HamiltonianCA}
A Courant algebroid $E$ with a connection $D$ and a section $\mu \in \Gamma(E^*)$ is called \textbf{Hamiltonian} 
if (H2) and (H3) are satisfied.
%If the condition is satisfied on a neighborhood of every point in $M$, 
%it is called locally Hamiltonian.
\end{definition}
We can add the following condition corresponding to (H1).
% in \cite{Blohmann:2018}. 
\medskip\\
\noindent
(H1):
$E$ is \emph{presymplectically anchored with respect to $D$} if
%\begin{eqnarray}
$D \gamma = 0$.
\medskip\\
\noindent
%\label{HH1}
%\end{eqnarray}
Here $D$ is a dual connection on $E^*$ defined by 
%\begin{eqnarray}
$d \bracket{\mu}{e} = \bracket{D\mu}{e} + \bracket{\mu}{De}$,
%\end{eqnarray}
for all sections $\mu \in \Gamma(E^*)$ and $e \in \Gamma(E)$.
%The dual connection extends to a degree 1 operator on $\Omega^k(M, E^*)$.

The condition (H1) is regarded as a flatness condition of the connection $D$ since $D \gamma = D^2 \mu$. In this paper, we do not assume the condition (H1) for the existence of a momentum section. (H1) can be imposed as an extra condition.

%%%%%%%%%%%%%%%%%%%%%%%%%%%%%%%%%%%%%%%%%%%%%%%%%%%%%%%%%%%%%%%%%%%%
%%%%%%%%%%%%%%%%%%%%%%%%%%%%%%%%%%%%%%%%%%%%%%%%%%%%%%%%%%%%%%%%%%%%
%%%%%%%%%%%%%%%%%%%%%%%%%%%%%%%%%%%%%%%%%%%%%%%%%%%%%%%%%%%%%%%%%%%%
\subsection{Examples}
\subsubsection{Trivial Lie algebra bundle: momentum map}
A momentum section is a generalization of a momentum map on a symplectic 
manifold with a Lie group action. Definitions of a momentum section 
(H2) and (H3) reduce to definitions of a momentum map if a Courant
algebroid $E$ is a trivial bundle of a Lie algebra with an inner product $\bracket{-}{-}$.

Suppose $B$ is nondegenerate, i.e., $B$ is a symplectic form. 
Consider an action Lie algebroid on the trivial bundle 
$E = M \times \mathfrak{g}$.
It means that an infinitesimal Lie algebra action is given by 
a bundle map $\rho: \mathfrak{g} \times M \rightarrow TM$, such that
\begin{eqnarray}
[\rho(e_1), \rho(e_2)] &=& \rho([e_1, e_2]).
\end{eqnarray}
The bracket in left hand side is a Lie bracket of vector fields.
%means that $\rho = \rho^i_a (x)e^a \partial_i$ is a Killing vector.
An action Lie algebroid is regraded as a special case of a Courant algebroid
if a Lie algebra has an inner product.
In this case, we can take a zero connection, $D = d$, and
axioms of a momentum section reduce to the following equations.
\medskip\\
\noindent
(H2)
A section $\mu \in \Gamma(M \times \mathfrak{g}^*)$ is regarded as a map
$\mu: M \rightarrow \mathfrak{g}^*$.
Then, Equation \eqref{HH2} is 
\begin{eqnarray}
\rd \mu(e) = \iota_{\rho(e)} B.
\label{momentmap2}
\end{eqnarray}
A map $\mu$ is a Hamiltonian for the vector field $\rho(e)$.
\medskip\\
\noindent
(H3)
We substitute Equation \eqref{momentmap2} to
the condition (H3), i.e., Equations \eqref{HH3} and \eqref{HH4}.
%, i.e. $\partial_i \mu_a = B_{ij} \rho^j_a$
\if0
\begin{eqnarray}
\rho^i_{[a} \partial_i \mu_{b]} - C_{ab}^c \mu_c
= - \gamma_{i[a} \rho^i_{b]}
= B_{ij} \rho^j_{a} \rho^i_{b},
\end{eqnarray}
is
\begin{eqnarray}
\rho^i_a \partial_i \mu_b - C_{ab}^c \mu_c = 0.
\end{eqnarray}
i.e.
\fi
Equation \eqref{HH4} is trivial, and
Equation \eqref{HH3} becomes
\begin{eqnarray}
\mathrm{ad}^*_{e_1} \mu(e_2) = \bracket{\mu}{[e_1, e_2]}.
\label{momentmap3}
\end{eqnarray}
for $e_1, e_2 \in \mathfrak{g}$.
This means that $\mu$ is $\mathfrak{g}$-equivariant.
Note that in this case, Equation \eqref{momentmap2} automatically leads 
(H1) 
%Equation \eqref{momentmap1} 
since $\rd^2 =0$:
\medskip\\
\noindent
(H1)
\begin{eqnarray}
{\rd \gamma
%\gamma 
= \rd^2 \mu= 0.}
\label{momentmap1}
\end{eqnarray}
%This means that $\rho(e)$ is a symplectic vector field.
%

Independent conditions are
\eqref{momentmap2} and \eqref{momentmap3}. It shows that
$\mu$ is an infinitesimally equivariant momentum map.

%Many examples of momentum sections which are not momentum maps have been discus%sed in \cite{Blohmann:2018}. One can refer to more examples.

%%%%%%%%%%%%%%%%%%%%%%%%%%%%%%%%%%%%%%%%%%%%%%%%%%%%%%%%%%%%%%%%%%%%
%%%%%%%%%%%%%%%%%%%%%%%%%%%%%%%%%%%%%%%%%%%%%%%%%%%%%%%%%%%%%%%%%%%%
%%%%%%%%%%%%%%%%%%%%%%%%%%%%%%%%%%%%%%%%%%%%%%%%%%%%%%%%%%%%%%%%%%%%
\subsubsection{Standard Courant algebroid}
We consider the standard Courant algebroid 
with $H$-flux in Example \ref{starndardCA}.
The homological vector field \eqref{homologicalvf}
is 
\begin{align}
Q &= \eta^i \frac{\partial}{\partial x^i}
+ \left(p_i + \frac{1}{2} h_{ijk} \eta^j \eta^k \right) 
\frac{\partial}{\partial \xi_k}
+ \frac{1}{3!} \partial_i h_{jkl} \eta^j \eta^k \eta^l
\frac{\partial}{\partial p_i},
\end{align}
where $\eta^a$ is decomposed to degree 1 coordinates $\eta^a = (\eta^i, \xi_i)$ of the fiber coordinates of $T[1]M$ and $T^*[1]M$,
and we can take the metric,
\begin{eqnarray}
  k = \left(
    \begin{array}{cc}
      0 & 1  \\
      1 & 0 
    \end{array}
  \right).
\label{fibermetric}
\end{eqnarray}
A momentum section $\mu \in \Gamma(T^*M \oplus TM)$ is decomposed to
$\mu = \nu + Z$, where $\nu \in \Omega(M)$ and
$Z \in \mathfrak{X}(M)$.

The condition (H2), Equation \eqref{HH2} is written as
\begin{eqnarray}
D Z = 0,
\\
D \nu = - B.
\end{eqnarray}
where $D$ is an affine connection on $TM$ and the corresponding induced connection on $T^*M$.
The condition (H3), Equation \eqref{HH3} reduces to 
\begin{eqnarray}
&& (\rd \nu + \iota_Z h)(X, Y) = - B(X, Y),
\end{eqnarray}
for all $X, Y \in \mathfrak{X}(M)$.

%%%%%%%%%%%%%%%%%%%%%%%%%%%%%%%%%%%%%%%%%%%%%%%%%%%%%%%%%%%%%%%%%%%%
%%%%%%%%%%%%%%%%%%%%%%%%%%%%%%%%%%%%%%%%%%%%%%%%%%%%%%%%%%%%%%%%%%%%
%%%%%%%%%%%%%%%%%%%%%%%%%%%%%%%%%%%%%%%%%%%%%%%%%%%%%%%%%%%%%%%%%%%%
\section{Constrained Hamiltonian mechanics}\label{constrainedHs}
We consider a nontrivial example of a momentum section of a Courant algebroid:
a constrained Hamiltonian mechanics system on the cotangent bundle $T^*M$ over 
a smooth manifold $M$ consistent with the Courant algebroid $E$.

Take local coordinates $(x^i, p_i)$ on the cotangent bundle $T^*M$.
The constraint Hamiltonian mechanics has two functions on a cotangent bundle (a phase space in physical terminology), a Hamiltonian $H$ and constraints $G_a$, which satisfy Poisson brackets,
\begin{eqnarray}
\{H, G_a\} &=& \sigma_a^b G_b,
\label{HGG} \\
\{G_a, G_b\} &=& C_{ab}^c G_c.
\label{GGG}
\end{eqnarray}
$H = H(x, p)$ is a function on $T^*M$ and 
$G_a = G_a(x, p)$ is a function on $T^*M$ taking a value on $E^*$.
$\sigma_a^b = \sigma_a^b(x, p)$ and $C_{ab}^c = C_{ab}^c(x, p)$ are local functions on the cotangent bundle.
The above equations \eqref{HGG} and \eqref{GGG}
are a realization of actions of groups or groupoids in the mechanics.
In this paper, we consider an action of the Courant algebroid, thus,
it is assumed that $G_a$ has the Courant algebroid structure.

%%%%%%%%%%%%%%%%%%%%%%%%%%%%%%%%%%%%%%%%%%%%%%%%%%%%%%%%%%%%%%%%%%%%
\subsection{Courant algebroid structure on constraints}\label{homogeneous}
First, in this section, we consider the simplest $G_a$ with a Courant algebroid structure.

A cotangent bundle has the canonical symplectic form,
\begin{eqnarray}
&& \omegac = \rd x^i \wedge \rd p_i,
\end{eqnarray}
which gives the canonical Poisson bracket,
\begin{eqnarray}
&& 
\{x^i, p_j\} = \delta^i_j.
\end{eqnarray}

Suppose that $G_a$ is \textit{linear} with respect to the momentum 
$p_i$. We can assume that
\begin{eqnarray}
G_a &=& \rho^i_a(x) p_i,
\label{constraint2}
\end{eqnarray}
where $\rho^i_a$ is a local function of $x$.
We denote this function by $\rho$ since later, we identify the function as
the local coordinate expression of the anchor map.
We impose the following Poisson brackets,
\begin{eqnarray}
\{G_a, G_b\} &=& C_{ab}^c G_c,
\label{constraintPoisson}
\end{eqnarray}
with some function $C_{ab}^c(x)$.
Equation \eqref{constraintPoisson} means 
that $G_a$ consists of a Lie algebra under the Poisson bracket, 
Then, $G_a$ is called first class constraints.
Note that from order counting of $p$, $C_{ab}^c(x)$ is a zeroth order with respect to $p$ and must only depend on $x$.
Referring to the identity $\rho([e_1, e_2]_D) = [\rho(e_1), \rho(e_2)]$
of the Courant algebroid
corresponding to Equation \eqref{CAidentity2}, 
the straightforward calculation gives
that Equation \eqref{constraintPoisson} is satisfied if a coefficient function $\rho^i_a(x)$ is the local coordinate expression of the anchor map $\rho$ and $C_{ab}^c = k^{cd} f_{cab}$.
Therefore, $G_a$ satisfies the following Poisson bracket,
\begin{eqnarray}
\{G_a, G_b\} &=& k^{cd} f_{cab} G_c.
\label{constraintPoisson2}
\end{eqnarray}

Another consistency check of this choice is needed.
The Jacobi identity $\{G_a, \{G_b, G_c\}\} + \mbox{cyclic}(abc) =0$ must be satisfied. The Jacobi identity is proved using identities
\eqref{CAidentity1}-- \eqref{CAidentity3} of the Courant algebroid.
Thus the Poisson algebra of $G_a$ is consistent with the Courant algebroid structure on $E$.

Note that the converse claim is not true.
In fact, suppose a function $G_a$ in \eqref{constraint2} 
without any assumption of a structure on the vector bundle $E$.
We only need one identity \eqref{CAidentity2} to satisfy 
Equation \eqref{constraintPoisson}.
The Jacobi identity 
$\{G_a, \{G_b, G_c\}\} + \mathrm{Cycl}(abc) =0$ impose
the equation for $\rho^i_a$, $k^{ab}$ and $f_{abc}$,
\begin{eqnarray}
\rho^i_{c} \partial_i f_{dab}+ k^{ef} f_{eab} f_{cdf} + 
(abc \ \mbox{cyclic})
= N_{abcd},
%= c \rho^i_d E_{iabc},
\end{eqnarray}
where $\partial_i = \tfrac{\partial}{\partial x^i}$, and the right hand side is an arbitrary $(abc)$-antisymmetric tensor $N_{abcd}(x)$ such that $\rho^j_e k^{ed} N_{abcd} =0$.
There are ambiguities to choose a tensor $N_{abcd}$ for the Jacobi identity.
The simplest solution is $N_{abcd}= 0$, which give a Lie algebroid structure
on the vector bundle $E$. This case was analyzed in \cite{Ikeda:2018rwe}. 
We took another solution,
%\eqref{supercondition01}--\eqref{supercondition04},
\begin{eqnarray}
N_{abcd}= \rho^i_d \partial_i f_{abc},
\end{eqnarray}
under a Courant algebroid structure on $E$ because we can prove that
\begin{eqnarray}
\rho^i_e k^{ed} N_{abcd}= \rho^i_e k^{ed} \rho^i_d \partial_i f_{abc} =0,
\end{eqnarray}
from the identity \eqref{CAidentity1} and \eqref{CAidentity3}.

There are possibilities of other solutions.
% such as higher Leibniz Lie $n$-algebroids.
The summary is as follows.
\begin{theorem}
Let $E$ be a Courant algebroid. Then, 
constraints \eqref{constraint2} satisfy the Poisson bracket
\eqref{constraintPoisson2} and the Jacobi identity.
\end{theorem}
$G_a = \rho^i_a(x) p_i$ is the first class constraint if $E$ is a Courant algebroid.

%%%%%%%%%%%%%%%%%%%%%%%%%%%%%%%%%%%%%%%%%%%%%%%%%%%%%%%%%%%%%%%%%%%%
%%%%%%%%%%%%%%%%%%%%%%%%%%%%%%%%%%%%%%%%%%%%%%%%%%%%%%%%%%%%%%%%%%%%
%%%%%%%%%%%%%%%%%%%%%%%%%%%%%%%%%%%%%%%%%%%%%%%%%%%%%%%%%%%%%%%%%%%%
\subsection{Free Hamiltonian}\label{free}
In this section, we introduce a Hamiltonian in the mechanics and discuss consistency with a Courant algebroid.

We consider the simple free Hamiltonian,
\begin{eqnarray}
H = \frac{1}{2} g^{ij}(x) p_i p_j,
\label{freeHamiltonian}
\end{eqnarray}
where $g^{ij}(x)$ is a symmetric tensor which is identified an inverse of a metric $g \in \Gamma(T^*M \otimes T^*M)$ on $M$. 
Suppose that the Hamiltonian has a symmetry generated by $G_a$,
i.e., the following Poisson bracket is imposed,
\begin{eqnarray}
\{H, G_a \} &=& \sigma_a^b G_b, 
\label{HGequation}
\end{eqnarray}
where $\sigma_a^b = \sigma_a^b(x, p)$ is some function of $x$ and $p$.
From the order counting of $p$, a function $\sigma_a^b$ must be a linear function of $p$.
Thus we can assume that
\begin{eqnarray}
\{H, G_a \} &=& - \Gamma_a^{bi} (x) p_i G_b.
\label{HGequationomega}
\end{eqnarray}
where $\Gamma_a^{bi}(x)$ is a function of $x$.
Equations \eqref{constraintPoisson} and \eqref{HGequationomega} must be 
covariant under transformations of the fiber of $E$ using a transition 
function $M_a^b$. 
This requires that $\Gamma_{ai}^b = g_{ij} \Gamma_a^{bj}$ transforms 
as a connection $1$-form on $E$,
\begin{eqnarray}
\Gamma' = M \Gamma M^{-1} + \rd M M^{-1}.
\end{eqnarray}

We denote the Courant algebroid connection defined by $\Gamma_{ai}^{b}$ 
by $D: \Gamma(E) \rightarrow \Gamma(E \otimes T^*M)$.
Substituting Equations \eqref{constraint2} and \eqref{freeHamiltonian} to 
\eqref{HGequationomega}, we obtain 
\if0
the following local identities,
\begin{eqnarray}
&& g^{ik} \partial_k \rho^j_a + g^{kj} \partial_k \rho^i_a
- \rho^k_a \partial_k g^{ij} = \Gamma_a^{bi} \rho^j_b + \Gamma_a^{bj} \rho^i_b.
\label{equationg}
\end{eqnarray}
Using the connection $D$,
Equation \eqref{equationg} is written as the coordinate independent form,
\fi
the condition for the metric $g$,
\begin{eqnarray}
&& {}^ED g = 0,
\label{diffmetric}
\end{eqnarray}
where ${}^ED$ is an $E$-connection ${}^ED: \Gamma(TM) \rightarrow \Gamma(TM \otimes E^*)$ on $TM$ defined by
\begin{eqnarray}
{}^ED_{\! e} v := L_{\rho(e)} v + \rho(D_v e),
\label{Econnection}
\end{eqnarray}
where $v \in \mathfrak{X}(M)$ is a vector field and $e \in \Gamma(E)$
is a section of $E$. $L$ is the Lie derivative.
An $E$-connection is extended to a covariant derivative on the tensor product space of $TM$ and $T^*M$ similar to a normal connection. Especially, ${}^ED$ in Equation \eqref{diffmetric} is the induced $E$-connection on the space $T^*M \otimes T^*M$.
In summary, we obtain the following theorem.
\begin{theorem}
Let $E$ be a Courant algebroid. Then, 
Equation \eqref{HGequation} is satisfied
if and only if Equation \eqref{diffmetric} is satisfied.
\end{theorem}
This theorem is the compatibility condition with a metric $g$ and the mechanics with a Courant algebroid structure.

%%%%%%%%%%%%%%%%%%%%%%%%%%%%%%%%%%%%%%%%%%%%%%%%%%%%%%%%%%%%%%%%%%%%
%%%%%%%%%%%%%%%%%%%%%%%%%%%%%%%%%%%%%%%%%%%%%%%%%%%%%%%%%%%%%%%%%%%%
%%%%%%%%%%%%%%%%%%%%%%%%%%%%%%%%%%%%%%%%%%%%%%%%%%%%%%%%%%%%%%%%%%%%
\subsection{Inhomogeneous generalization and momentum section}
\label{inhomomecha}
We generalize the mechanics in the previous section to the system with constraints and Hamiltonian inhomogeneous with respect to the order of $p$. A momentum section appears in the mechanics.

We generalize the constraint $G_a$ in addition to the zeroth order term of $p$,
\begin{eqnarray}
G_a &=& \rho^i_a(x) p_i + \alpha_a(x),
\label{inhomoGs}
\end{eqnarray}
where $\alpha_a$ is a local section of $E^*$.
The Hamiltonian is also generalized to inhomogeneous one as follows,
\begin{eqnarray}
H = \frac{1}{2} g^{ij}(x) p_i p_j + \beta^i(x) p_i + V(x),
\label{inhomoHs}
\end{eqnarray}
where $\beta = \beta^i \partial_i \in \mathfrak{X}(M)$ is a vector field on 
$M$ and $V \in C^{\infty}(M)$.

The second term $\beta^i(x) p_i$ in the Hamiltonian is absorbed by 
redefining the conjugate momentum $p_i$ as 
$p'_i = p_i + A_i$, where $A_i = g_{ij} \beta^j$ is a 1-form.
The 1st order term of $p$ in the Hamiltonian is absorbed to other terms as
expected,
\begin{eqnarray}
H = \frac{1}{2} g^{ij}(x) p'_i p'_j + V'(x),
\label{inhomoH}
\end{eqnarray}
where
\begin{eqnarray}
V'(x) = V(x) - \frac{1}{2} g^{ij} A_i A_j.
\end{eqnarray}
The symplectic form is changed to
\begin{eqnarray}
&& \omega = \rd p_i' \wedge \rd x^i + \frac{1}{2} B_{ij}(x) \rd x^i \wedge \rd x^j,
\end{eqnarray}
with $B= dA \in \Omega^2(M)$.
Obviously, $B$ is closed, thus it defines a pre-symplectic form on $M$.
The Poisson bracket of $p$'s becomes
\begin{eqnarray}
&& \{p'_i, p'_j \} = - B_{ij}.
\end{eqnarray}
Using $p'$, $G_a$ is written as
\begin{eqnarray}
G_a = \rho^i_a(x) p'_i + \mu_a,
\label{inhomoG}
\end{eqnarray}
where $\mu_a = \alpha_a - \rho^i_a A_i$ is a section of $E^*$.

In order to construct consistent constrained mechanics, we require
that $G_a$ satisfies the same equation as
Equation \eqref{constraintPoisson},
\begin{eqnarray}
\{G_a, G_b\} &=& k^{cd} f_{cab} G_d.
\label{GGpoisson2}
\end{eqnarray}
which means that $G_a$ is the first class constraint.
This condition imposes the following conditions to $\mu$ 
\begin{eqnarray}
{}^E \rd \mu(e_1, e_2) = \rho^*(B)(e_1, e_2),
\label{conditionH3}
\end{eqnarray}
for $e_i \in \Gamma(E)$
in addition to the Courant algebroid structure in subsection
\ref{homogeneous}.
%Here ${}^E \rd$ is the Courant algebroid differential on 
%$\Gamma(\wedge^{\bullet} E^*)$, which is equivalent to the homological 
%vector field $Q$ on $T^*[2]E[1]$.
Here $\rho^*$ is the induced map of the anchor to $\Omega^{\bullet}(M)$, 
mapping ordinary differential forms to $E$-differential forms,
$\rho^*(B)= \frac{1}{2} B_{ij} \rho_a^i \rho_b^j q^a q^b \in \Gamma(\wedge^2E^*)
%\equiv{}^E\Omega^2(M)
$. 
If we define $\gamma \in \Omega^1(M, E^*)$ as
\begin{eqnarray}
\bracket{\gamma(v)}{e} = - B(v, \rho(e)),
\label{conditionH2}
\end{eqnarray}
for all $v \in \mathfrak{X}(M)$ and $e \in \Gamma(E)$,
Equation \eqref{conditionH3} is written as
\begin{eqnarray}
&& {}^E \rd \mu(e_1, e_2) = - \bracket{\gamma(\rho(e_1))}{e_2},
\label{conditionH32}
\end{eqnarray}
which is the same equation as \eqref{HH3}, i.e., (H3) in the definition of the momentum section.
Equation \eqref{HH4} is automatically satisfied.
%

\if0
Equation \eqref{conditionH3}
includes two independent condition
\begin{eqnarray}
&& {}^E\nabla \mu(e_1, e_2) = \rho^*(B)(e_1, e_2),
\label{conditionH3}
\\
&& \bracket{\mu(e_1)}{\rho(e_2) f} =0, 
\end{eqnarray}
for all $e_1, e_2 \in \Gamma(E)$ and $f \in C^{\infty}(M)$.
Here
\begin{eqnarray}
({}^E\nabla \mu)_{ab} &=& \frac{1}{2} (\rho^i_a \partial_i \mu_b
- \rho^i_a \partial_i \mu_b + k^{cd} f_{abc} \mu_d).
\end{eqnarray}
\fi

Thus, we obtain the following result.
\begin{theorem}
Assume $G_a$ in Equation \eqref{inhomoG} under a Courant algebroid structure. Then Equation \eqref{GGpoisson2} gives the identity \eqref{conditionH32}, i.e. the condition (H3).
\end{theorem}

Next, we require the same equation as Equation \eqref{HGequation}
for the Poisson bracket of $H$ and $G_a$,
\begin{eqnarray}
\{H, G_a \} &=& \sigma_a^b G_b,
\label{HGequationinhomo}
\end{eqnarray}
i.e., $G_a$ is required to generate a symmetry of the Hamiltonian.
Here the coefficient function is generalized to an inhomogeneous function $\sigma_a^b(x) = - \Gamma_a^{bi}(x) p_i - \tau_a^b(x)$,
where $\tau$ is a local endomorphism of sections of $E$.
%From the consistency of Equation \eqref{HGequationinhomo} under the 
%given by equivalence of $G_a$,
If we require $G_a$ is a global section of $E^*$, 
a gauge transformation must transform $\Gamma$ and $\tau$ as
 \begin{eqnarray}
 \Gamma^{\prime} &=& M \Gamma M^{-1} + d M M^{-1},
\label{transformationofomega}
\\
 \tau^{\prime} &=& M \tau M^{-1} + \iota_{\beta} (d M M^{-1}),
\label{transformationoftau}
 \end{eqnarray}
where $M^a_b$ is a transition function.
Thus, $\Gamma_a^b = \Gamma_{aj}^{b}dx^j$ transform as a connection 1-form 
on $E$,
%and
%$g(-,-)$ is an inner product defined by the metric $g$. 
and $\tau = \iota_{\beta} \Gamma  = g(\Gamma, A)$.
Thus 
\begin{eqnarray}
\{H, G_a \} &=& 
%- g^{ij} \Gamma_{aj}^b (p_i + A_j) G_b = 
- g^{ij} \Gamma_{aj}^b p_i' G_b,
\label{HGequationinhomo}
\end{eqnarray}

We compute identities for coefficient functions by substituting 
concrete expressions of $H$, $G_a$ and $\sigma$ 
\eqref{inhomoGs}, \eqref{inhomoHs} and \eqref{HGequationinhomo}
to Equation \eqref{HGequationinhomo}.
The $p'^2$ order of Equation \eqref{HGequationinhomo} gives Equation \eqref{diffmetric}, which is the same consistency condition for the metric in the homogeneous case.
The $p'^1$ order of Equation \eqref{HGequationinhomo} gives
a new condition for $\mu$,
\begin{eqnarray}
D \mu = \gamma,
\label{identityp1}
\end{eqnarray}
which is the condition (H2) in the definition of momentum section.
\if0
\begin{eqnarray}
\partial_i \mu_a - \Gamma_{ai}^b \mu_b &=& - \rho^k_a B_{ik}.
\end{eqnarray}
\fi
The $0$-th order of $p$ gives a condition for the potential function $V'$
as
\begin{eqnarray}
{}^E \rd V' = 0,
\label{identityp0}
\end{eqnarray}
which is independent of other data including the momentum section.

\begin{theorem}\label{inhomoH2}
Assume inhomogeneous $G_a$ and $H$ in Equation \eqref{inhomoGs} and \eqref{inhomoHs} under a Courant algebroid structure. Then Equation 
\eqref{HGequationinhomo} gives
the identities \eqref{diffmetric}, \eqref{identityp1} and \eqref{identityp0}.
Especially, the Poisson bracket \eqref{HGequationinhomo} gives the condition (H2).
\end{theorem}

We summarize all results of the constrained Hamiltonian system in this section.
\begin{proposition}\label{inhomoresult}
Let $E$ be a Courant algebroid over $M$.
The inhomogeneous constrained Hamiltonian system on $T^*M$ 
with the constraint \eqref{inhomoGs} and the Hamiltonian \eqref{inhomoHs}.
If we require consistency with a Courant algebroid $E$ over $M$,
we obtain Poisson brackets \eqref{GGpoisson2} and \eqref{HGequationinhomo}
with a connection 1-form $\Gamma_a^b$,
and
\begin{eqnarray}
&& {}^E Dg =0,
\\
&& \tau = g(\Gamma, A),
\\
&& {}^E \rd V' = 0,
\end{eqnarray}
Moreover, $\mu = \alpha - \iota_{\rho} A$ is a bracket compatible $D$-momentum section with respect to the pre-symplectic structure defined by $B =dA$.
Here $A = \iota_{\beta} g \in \Omega^1(M)$ and $D$ is a connection induced from $\Gamma_a^b$.
\end{proposition}

%%%%%%%%%%%%%%%%%%%%%%%%%%%%%%%%%%%%%%%%%%%%%%%%%%%%%%%%%%%%%%%%%%%%%
%%%%%%%%%%%%%%%%%%%%%%%%%%%%%%   SEC      %%%%%%%%%%%%%%%%%%%%%%%%%%
%%%%%%%%%%%%%%%%%%%%%%%%%%%%%%%%%%%%%%%%%%%%%%%%%%%%%%%%%%%%%%%%%%%%%
\section{BFV and BV description: BFV formalism}\label{BFVBV}
In order to formulate cohomological description of the momentum section and 
the Hamiltonian Courant algebroid,
we consider the BFV and BV formalism 
\cite{Batalin:1977pb, Batalin:1983pz, Batalin:1981jr, Batalin:1984jr}
of the constrained Hamiltonian mechanics
in Section \ref{constrainedHs}.
In the Lie algebroid case, the BFV and BV formalism were analyzed in the paper \cite{Ikeda:2018rwe}.

In this section, we consider the BFV formalism.
The classical BFV formalism is defined on a graded manifold $T^*[2]E[1]$, where $T^*[2]M$ is an original phase space of the Hamiltonian mechanics and $E[1]$ 
is a vector bundle whose fiber degree is shifted by 1.
The BFV data consist of even and odd functions $H_{BFV}$ and $S_{BFV}$ satisfying
\begin{eqnarray}
&& \{S_{BFV}, S_{BFV} \}=0,
\label{BFVequation1}
\\
&& \{S_{BFV}, H_{BFV} \}=0,
\label{BFVequation2}
\\
&& \{H_{BFV}, H_{BFV} \}=0,
\label{BFVequation3}
\end{eqnarray}
where $\{-,-\}$ is a Poisson bracket.
Note that Equation \eqref{BFVequation1} is not a trivial equation since $S_{BFV}$ is an odd function though Equation \eqref{BFVequation3} is trivial.
Two functions are constructed to be equivalent to the original Hamiltonian 
mechanics.

%%%%%%%%%%%%%%%%%%%%%%%%%%%%%%%%%%%%%%%%%%%%%%%%%%%%%%%%%%%%%%%%%%%%%
%%%%%%%%%%%%%%%%%%%%%%%%%%%%%%   SEC      %%%%%%%%%%%%%%%%%%%%%%%%%%
%%%%%%%%%%%%%%%%%%%%%%%%%%%%%%%%%%%%%%%%%%%%%%%%%%%%%%%%%%%%%%%%%%%%%
\subsection{BFV of homogeneous constrained Hamiltonian mechanics}\label{BFVhomo}
We consider the homogeneous constrained Hamiltonian mechanics in Section \ref{homogeneous} and \ref{free}.

We introduce odd coordinates $\eta^a$ of degree one on the fiber of $E[1]$
in addition to local coordinates $(x^i, p_i)$ of the cotangent bundle $T^*M$.
Degree of $(x^i, p_i)$ is assigned by $(0, 2)$.
The dual bundle $E^*[1]$ is identified $E[1]$ by using the fiber metric $\bracket{-}{-}$.
Since $T^*[2]E[1]$ is the cotangent bundle, the canonical graded symplectic form is introduced,
\begin{eqnarray} \label{BFVsymplectic}
\omega_{BFV} = \rd x^i \wedge \rd p_i + \frac{1}{2} k_{ab} \rd \eta^a \wedge \rd \eta^b,
\end{eqnarray}
as an extension of the canonical symplectic form in the phase space of the mechanics.
The canonical Poisson brackets are
\begin{eqnarray}
&& \{x^i, p_j \} = \delta^i_j,
\\ 
&& \{\eta^a, \eta^b \} = k^{ab},
\end{eqnarray}
and other Poisson brackets vanish.
The BFV charge $S_{BFV}$ is nothing but the Chevalley-Eilenberg operator 
of the Courant algebroid complex, which is the Hamiltonian function of 
the homological function $Q$ in Equation \eqref{homologicalvf}.
The formula gives
\begin{eqnarray} \label{SBFV}
S_{BFV} = \eta^a G_a - \frac{1}{3!} f_{abc} \eta^a \eta^b \eta^c,
\label{SBFVhomo}
\end{eqnarray}
where the constraint $G_a$ is given in \eqref{constraintPoisson2}.
We can easily check that $S_{BFV}$ satisfies ${}^E \rd= Q= \{S_{BFV}, -\}$,
and Equation \eqref{BFVequation1}, which is equivalent to $Q^2=0$.
Since $G_a$ is homogeneous with respect to $p_i$ and a degree 2 function, 
$S_{BFV}$ is a homogeneous function of degree 3.

Next, we determine the BFV Hamiltonian $H_{BFV}$ by solving
Equation \eqref{BFVequation2}.
$H_{BFV}$ is expanded by $\eta^a$ as
\begin{eqnarray} \label{HBFV}
H_{BFV} = \sum_{I=0}^{\infty} H^{(I)}_{BFV},
\end{eqnarray}
where $H^{(I)}_{BFV}$ is the $2I$-th order term,
Each term is even order of $\eta^a$ since $H_{BFV}$ is an even function.
\begin{eqnarray}
H^{(I)}_{BFV} = \frac{1}{(2I)!} 
\eta^{a_1} \ldots \eta^{a_{2I}} H^{(I)}_{BFV a_1 \cdots a_{2I}}(x, p).
\label{HBFVI}
\end{eqnarray}
The $0$-th part is fixed to the classical Hamiltonian $H^{(0)}_{BFV} = H$.
Higher terms are determined recursively by substituting \eqref{SBFV}, \eqref{HBFV} and \eqref{HBFVI} to Equation \eqref{BFVequation2}, i.e., $\{S_{BFV}, H_{BFV} \}=0$.

%We assign degree of each coordinate as
%$\mathrm{deg}\, x^i =0$, $\mathrm{deg}\, \eta^a =1$, $\mathrm{deg}\, p_i =2$.
%Then, $\mathrm{deg}\, \omega = 2$, $\mathrm{deg}\, H = 4$, 
%$\mathrm{deg}\, S_{BFV} = 3$.
Since degree of the classical Hamiltonian $H$ is 4, the BFV Hamiltonian $H_{BFV}$ must be a degree 4 function from homogeneity.
Thus, the BFV Hamiltonian is expanded as
\begin{eqnarray}
H_{BFV} &=& H^{(0)}_{BFV} + H^{(1)}_{BFV} + H^{(2)}_{BFV},
\\
H^{(0)}_{BFV} &=& H_{cl}(x, p) = \frac{1}{2} g^{ij}(x) p_i p_j,
\\
H^{(1)}_{BFV} &=& \frac{1}{2} H^{(1)}{}^{i}_{ab}(x) p_i \eta^a \eta^b,
\\
H^{(2)}_{BFV} &=& \frac{1}{4!} H^{(2)}_{abcd}(x) \eta^a \eta^b \eta^c \eta^d,
\\
H^{(I)}_{BFV} &=& 0, \qquad\qquad \mbox{for}\ I \geq 3,
\end{eqnarray}
Substituting this expression to \eqref{BFVequation2}, we obtain 
%$\{S_{BFV}, H_{BFV} \}=0$ gives the solution of the second order term,
\begin{eqnarray}
H^{(1)}_{BFV} &=& 
%\frac{1}{2} \sigma_{a}^c(x, p) k_{cb} \eta^{a} \eta^{b} =
\frac{1}{2} \eta^{a} \eta^{b} g^{ij} \Gamma_{abi} p_j,
\end{eqnarray}
which depends on the connection $\Gamma$, where
$\Gamma_{abj} \equiv \Gamma_{aj}^c k_{cb}$.

In order to solve $H^{(2)}_{BFV}$,
we covariantize the conjugate momentum $p_i$ as 
\begin{eqnarray}
&& p_i^{\nabla} = p_i + \frac{1}{2} \Gamma_{abi} \eta^a \eta^b.
\end{eqnarray}
In fact, $p_i^{\nabla}$ transforms covariantly under the super diffeomorphism on $E[1]$.
Poisson brackets of covariantized canonical coordinates are
\begin{eqnarray}
&& \{x^i, p_j^{\nabla} \} = \delta^i_j,
\\
&& \{p_i^{\nabla}, \eta^a \} = \Gamma_{bi}^a \eta^b,
%- k^{ac} \Gamma_{cbi} \eta^b,
\label{Poissonpeta}
\\
&& \{\eta^a, \eta^b \} = k^{ab},
\\
&& \{p_i^{\nabla}, p_j^{\nabla} \} = - \frac{1}{2} R_{ija}^b k_{bc} \eta^a \eta^c,
\end{eqnarray}
where $R_{ija}^b$ is the curvature with respect to $\Gamma_{ai}^b$.
See Appendix \ref{RTS} for the concrete definition.
If we introduce the basis of $E$, $e_a$, it is natural 
that the Poisson bracket is defined as
\begin{eqnarray}
&& \{p_i^{\nabla}, e_a \} := - \Gamma_{ai}^b e_b,
\label{Poissonpe}
\end{eqnarray}
from $p_i^{\nabla}$ is the Hamiltonian for $-D_i$ and
the definition of a connection,
\begin{eqnarray}
&& D_i e_a = \Gamma_{ai}^b e_b,
\label{Poissonpe}
\end{eqnarray}
Considering $\eta = \eta^a e_a$, 
the Poisson bracket \eqref{Poissonpeta} is written in the covariant form,
\begin{eqnarray}
&& \{p_i^{\nabla}, \eta \} = 0.
\end{eqnarray}

The BFV charge function is also covariantized as
\begin{eqnarray}
S_{BFV} = \eta^a \rho_a^i p_i^{\nabla} 
- \frac{1}{3!} T_{abc} \eta^a \eta^b \eta^c,
\end{eqnarray}
where
\begin{eqnarray}
T_{abc} &=& f_{abc} - \rho^i_{[a} \Gamma_{bc]i},
\end{eqnarray}
is the $E$-torsion (the Gualtieri torsion) on the Courant algebroid.
An $E$-torsion $T \in \Gamma(\wedge^3 E^*)$ on a Courant algebroid $E$ 
is defined by \cite{Gualtieri:2003dx}
\begin{eqnarray}
T(e_1, e_2, e_3) &:=& - \frac{1}{2} 
\bracket{{}^ED_{e_1} e_2 - {}^ED_{e_2} e_1}{e_3}
+ \frac{1}{3} \bracket{[e_1, e_2]_C}{e_3} + (123\ \mbox{cyclic}),
\end{eqnarray}
for $e_I \in \Gamma(E)$.
Using the covariant coordinate $p_i^{\nabla}$, $H^{(1)}$ is absorbed to $H^{(0)}$. The BFV Hamiltonian is simplified to
\begin{eqnarray}
H_{BFV} &=& H^{\nabla(0)} + H^{\nabla(2)},
\label{covariantHBFV}
\end{eqnarray}
where
\begin{eqnarray}
H^{\nabla(0)} &=& \frac{1}{2} g^{ij} p_i^{\nabla} p_j^{\nabla},
\\
H^{\nabla(2)} &=& \frac{1}{4!} U_{abcd}(x) \eta^a \eta^b \eta^c \eta^d.
\end{eqnarray}
Here $U \in \Gamma(\wedge^4 E^*)$ is an $E$-4-form on $M$.
Substituting Equation \eqref{covariantHBFV} to \eqref{BFVequation2}
we obtain Equation of $U$,
%
\if0
\begin{eqnarray}
\{S_{BFV}, H^{\nabla(0)} \} &=& 
\frac{1}{3!} g^{jk} S_{jab}^c k_{cd} \eta^a \eta^b \eta^d p_k^{\nabla},
\\
\{S_{BFV}, H^{\nabla(2)} \} &=& 
\frac{1}{3!} \rho^i_d k^{de} U_{eabc} \eta^a \eta^b \eta^c p_i^{\nabla}
- \frac{1}{5!} (\rho^i_{[a} \nabla_i U_{bcde]}
+ k^{fg} T_{f[ab} U_{cde]g})
\eta^a \eta^b \eta^c \eta^d \eta^e,
\end{eqnarray}
%
%$U$ is a solution of two equations,
\begin{eqnarray}
&& \rho^i_d k^{de} U_{eabc} = g^{ij} S_{jab}^d k_{dc},
\label{equationU1}
\\
&& \rho^i_{[a} D_i U_{bcde]} + k^{fg} T_{f[ab} U_{cde]g} =0.
\label{equationU2}
\end{eqnarray}
\fi
\begin{eqnarray}
&& \bracket{\rho v}{U}(e_1, e_2, e_3) = \bracket{g^{-1}(S(e_1,e_2), v)}{e_3},
\label{equationU11}
\\
&& {}^E D U + \bracket{T}{U} =0,
\label{equationU12}
\end{eqnarray}
where $S_{jab}^c$ is the \textit{basic curvature} $S \in \Gamma(T^*M \otimes E \otimes \wedge^2 E^*)$ defined by 
\begin{eqnarray}
S = D T + 2 \mathrm{Alt}(\iota_{\rho} R),
\end{eqnarray}
where $\mathrm{Alt}$ denotes an antisymmetrization over $E^* \otimes E^*$
.
The second term is a kind of Bianchi identity of $U$.
If the basic curvature $S$ of $E$ is zero,
we can take the simplest solution $U=0$.
Even if $S \neq 0$,
we can obtain solutions of the BFV Hamiltonian $H_{BFV}$ by solving Equations
\eqref{equationU1} and \eqref{equationU2} under the geometric condition.
We summarize the result.
\begin{theorem}
There exists a solution of the BFV equations satisfying \eqref{BFVequation1}--\eqref{BFVequation3} if there exists $U \in \Gamma(\wedge^4 E^*)$ satisfying
\begin{eqnarray}
&& \bracket{\rho v}{U}(e_1, e_2, e_3) = \bracket{g^{-1}(S(e_1,e_2), v)}{e_3},
\label{equationU11}
\\
&& {}^E D U + \bracket{T}{U} =0,
\label{equationU12}
\end{eqnarray}
where $v \in \mathfrak{X}(M)$ and $e_I \in \Gamma(E)$.
\end{theorem}
We call a Courant algebroid $E$ with connection $D$ a \textit{Cartan Courant algebroid} if its basic curvature $S=0$ 
as in the case of a Lie algebroid \cite{Blaom}.

\begin{corollary}
There exists a solution of the BFV equations satisfying \eqref{BFVequation1}--\eqref{BFVequation3} if a Courant algebroid is the Cartan Courant algebroid.
\end{corollary}

%%%%%%%%%%%%%%%%%%%%%%%%%%%%%%%%%%%%%%%%%%%%%%%%%%%%%%%%%%%%%%%%%%%%%
%%%%%%%%%%%%%%%%%%%%%%%%%%%%%%   SEC      %%%%%%%%%%%%%%%%%%%%%%%%%%
%%%%%%%%%%%%%%%%%%%%%%%%%%%%%%%%%%%%%%%%%%%%%%%%%%%%%%%%%%%%%%%%%%%%%
\subsection{BFV in inhomogeneous system and momentum section}\label{BFVinhomo}
We consider a generalization to inhomogeneous constraints and Hamiltonian 
in Section \ref{inhomomecha}.

First we replace $p_i$ to $p_i = p_i' - A_i$ similar to the change of basis 
in Section \ref{inhomomecha}.
Then, the BFV symplectic form is changed to
\begin{eqnarray}
\omega_{BFV} = \rd x^i \wedge \rd p_i' + \frac{1}{2} B_{ij} \rd x^i \wedge \rd x^j + \frac{1}{2} k_{ab} \rd \eta^a \wedge \rd \eta^b.
\label{inhomoBFV}
\end{eqnarray}
The Poisson bracket of $\{p_i', p_j' \}$ is deformed as
\begin{eqnarray}
&& \{p_i', p_j' \} = - B_{ij}.
\end{eqnarray}
The BFV charge $S_{BFV}$ is formally the same equation \eqref{SBFVhomo}
as in the homogeneous case, 
\begin{eqnarray}
S_{BFV} = \eta^a G_a - \frac{1}{3!} f_{abc} \eta^a \eta^b \eta^c,
\label{SBFVinhomo}
\end{eqnarray}
since the Poisson bracket of $G_a$ is the same equation 
\eqref{HGequationinhomo}. However, the constraint $G_a$ is given by inhomogeneous Equation \eqref{inhomoG} including a momentum section $\mu$,
\begin{eqnarray}
G_a = \rho^i_a(x) p_i' + \mu_a.
\end{eqnarray}
Thus the covariantized BFV function is deformed by the momentum section $\mu$,
\begin{eqnarray}
S_{BFV} 
= \rho_a^i \eta^a p'{}_i^{\nabla} 
- \frac{1}{3!} T_{abc} \eta^a \eta^b \eta^c + \mu_a \eta^a.
\label{SBFVinhomocov}
\end{eqnarray}
%In fact, Equation \eqref{SBFVinhomo} satisfy $\{S_{BFV}, S_{BFV} \}=0$.
%From the dicussion in Section \ref{inhomomecha},
Straight calculation shows that under the Courant algebroid structure,
The second order term in $\{S_{BFV}, S_{BFV} \}=0$ is equivalent to the condition (H3) in the definition of the momentum section.
We expand the constraint $G_a = G_{2a} + G_{0a}$ by degree,
\begin{eqnarray}
G_{2a} &=& \rho^i_a(x) p_i',
\qquad
G_{0a} = \mu_a.
\end{eqnarray}
The corresponding expansion of $S_{BFV}$ by degree is
\begin{eqnarray}
S_{BFV3} &=& \eta^a G_{2a} - \frac{1}{3!} f_{abc} \eta^a \eta^b \eta^c,
\\
S_{BFV1} &=& \eta^a G_{0a}.
\end{eqnarray}
$S_{BFV1}$ is the term of the momentum section.
In the term of BFV formulation, the condition (H3) is written as
\begin{eqnarray}
\{S_{BFV3}, S_{BFV1}\} &=& - \frac{1}{2} \{S_{BFV3}, S_{BFV3}\}.
\label{momsec1}
\end{eqnarray}

The BFV Hamiltonian $H_{BFV}$ is also inhomogeneous since
the classical Hamiltonian $H_{cl}$ is
%$H^{(0)}_{BFV}$ is deformed to 
Equation \eqref{inhomoH},
\begin{eqnarray}
H^{(0)}_{BFV} &=& H_{cl} = \frac{1}{2} g^{ij}(x) p_i' p_j' + V'(x).
\end{eqnarray}
\if0
As in homogeneous case, we compute $H_{BFV}$ step by step by expanding 
the order of $\eta^a$,
\begin{eqnarray} \label{HBFV}
H_{BFV} = \sum_{I=0}^{\infty} H^{(I)}_{BFV},
= \sum_{I=0}^{\infty} \frac{1}{(2I)!} 
\eta^{a_1} \ldots \eta^{a_{2I}} H^{(I)}_{BFV a_1 \cdots a_{2I}}(x, p').
\label{HBFVI}
\end{eqnarray}
\fi
The solution of Equation $\{S_{BFV}, H_{BFV}\} =0$ is 
obtained by deforming $p_i$ to $p_i'$ in the solution $H_{BFV}$ for the homogeneous case.
We fix the deformed inhomogeneous BFV Hamiltonian as
\begin{eqnarray}
H_{BFV} &=& H^{\nabla(0)} + V'(x) + H^{\nabla(2)},
\label{inhomoHBFV}
\end{eqnarray}
where
\begin{eqnarray}
H^{\nabla(0)} &=& \frac{1}{2} g^{ij} p_i'{}^{\nabla} p_j'{}^{\nabla},
\\
H^{\nabla(2)} &=& \frac{1}{4!} U_{abcd}(x) \eta^a \eta^b \eta^c \eta^d.
\end{eqnarray}
Substituting inhomogeneous BFV functional \eqref{SBFVinhomo} and Hamiltonian \eqref{inhomoHBFV} to Equation \eqref{BFVequation2}, we obtain conditions.

$U$ is independent of a momentum section $\mu$ and additional inhomogeneous terms. The condition for $U$ is the same as homogeneous Hamiltonian case.
\begin{theorem}\label{solutionU}
There exists a solution of the BFV equations in inhomogeneous constrained Hamiltonian mechanics satisfying \eqref{BFVequation1}--\eqref{BFVequation3} if 
there exists $U \in \Gamma(\wedge^4 E^*)$ satisfying
\begin{eqnarray}
&& \bracket{\rho v}{U}(e_1, e_2, e_3) = \bracket{g^{-1}(S(e_1,e_2), v)}{e_3},
\label{equationU1}
\\
&& {}^E D U + \bracket{T}{U} =0,
\label{equationU2}
\end{eqnarray}
where $v \in \mathfrak{X}(M)$ and $e_I \in \Gamma(E)$.
\end{theorem}
Suppose a solution $U$. 
For other geometric quantities, we obtain the equivalent solution as in Theorem \ref{inhomoH2}
i.e., 
${}^E Dg =0$, ${}^E \rd V' = 0$ and $\mu = \alpha - \iota_{\rho} A$ satisfies
the condition (H2).

We can summarize the BFV formalism.
\begin{theorem}
Suppose a solution $U$ in Theorem \ref{solutionU}
and ${}^E D g=0$ and ${}^E \rd V'=0$.
Then, the BFV formalism with $S_{BFV}$ and $H_{BFV}$
of the inhomogeneous constrained Hamiltonian mechanics
is equivalent to the conditions (H2) and (H3) of a momentum section
on a Courant algebroid.
\end{theorem}
\if0
If the basic curvature $S=0$, there is the solution $U=0$
in Equations \eqref{equationU1} and \eqref{equationU2}.
Therefore we obtain,
\begin{corollary}
If the basic curvature $S=0$, 
there always exists the BFV formalism for a momentum section on a Courant algebroid.
\end{corollary}
\fi
We expand $H_{BFV}$ by degree,
\begin{eqnarray}
H_{BFV} &=& H_{BFV4} + H_{BFV0},
\\
H_{BFV4} &=& H^{\nabla(0)} + H^{\nabla(2)},
\\
H_{BFV4} &=& V'(x).
\end{eqnarray}
From the expansion of $S_{BFV}$ and $H_{BFV}$ with respect to degree,
The condition (h2) of the momentum section is written as
\begin{eqnarray}
\{S_{BFV1}, H_{BFV4} \} &=& \{S_{BFV3}, H_{BFV4} \}.
\label{momsec2}
\end{eqnarray}
%which is equivalent to the condition (H2) 
%under the assumption for the metric $g$ \eqref{diffmetric}.
% and $V'$ \eqref{identityp0}.

%%%%%%%%%%%%%%%%%%%%%%%%%%%%%%%%%%%%%%%%%%%%%%%%%%%%%%%%%%%%%%%%%%%%%
%%%%%%%%%%%%%%%%%%%%%%%%%%%%%%   SEC      %%%%%%%%%%%%%%%%%%%%%%%%%%
%%%%%%%%%%%%%%%%%%%%%%%%%%%%%%%%%%%%%%%%%%%%%%%%%%%%%%%%%%%%%%%%%%%%%
\section{BV-FHGD formalism}\label{BV}
The classical BV formalism is equivalent to a Lagrangian mechanics.
The BV formalism is defined on a graded manifold
$\calM_{BV} = T^*[-1] \calM_{BRST}$,
and consists of a nondegenerate odd Poisson bracket $(-,-)$ called 
the BV bracket and one even function $S_{BV}$ satisfying
\begin{eqnarray}
&& (S_{BV}, S_{BV}) =0.
\end{eqnarray}
The BV bracket is defined by an odd symplectic form $\omega_{BV}$.

An action functional in the Lagrangian formalism corresponding to 
the constrained Hamiltonian mechanics 
\eqref{constraintPoisson2} and \eqref{HGequation}
is 
\cite{Fisch:1989rm, Dresse:1990ba}
%\footnote{See also \cite {Ikeda:2020eft} about ambiguity of trivail gauge transformations.}
\begin{eqnarray}
S_{cl} = \int_{\bR} dt (p_i \dot{x}^i - H + \lambda^a G_a),
\label{classicalaction}
\end{eqnarray}
where $\dot{x}$ is the time derivative of $x$ and $\lambda^a$ is a Lagrange multiplier.
In fact, the Legendre transformation of canonical conjugates gives the Hamiltonian $H$ and the variation with respect to $\lambda^a$ gives the constraint equation $G_a \approx 0$.

Though there is a normal procedure to construct a BV bracket and a BV action functional \cite{Henneaux:1992ig}, in this paper, we use the Fisch-Henneaux-Grigoriev-Damgaard (FHGD) method \cite{Henneaux:1992ig, Grigoriev:1999qz, Grigoriev:2010ic} since the procedure is simpler than the normal BV construction.
The FHGD method is the formula to construct the BV symplectic form and 
the BV functional from the BFV data. See also \cite{Ikeda:2020eft}.

In this section, the method and formulas are briefly explained.

Let $z^I(t)$ denote fields in the BFV formalism including the ghosts. The BFV data are then given by a BFV symplectic form $\omega_{BFV}$, an odd function $S_{BFV}(\sigma)$, and an even function $H_{BFV}$.
The BFV symplectic form can be written as 
\begin{eqnarray}
\omega_{BFV} = \omega_{IJ}(z) \rd z^I \wedge \rd z^J,
\label{BFVsymplecticFHGD}
\end{eqnarray}
where $\omega_{IJ}(z)$ is a nondegenerate, graded-antisymmetric matrix.
Likewise, the BFV functional $S_{BFV}(z)$ and the BFV Hamiltonian $H_{BFV}(z)$ are functions of $z$.

The FHGD procedure in the formulation of Grigoriev and Damgaard, then works as follows. First, for each coordinate $z^I(t)$ one introduces a superpartner field $w^I(t)$. 
In addition, it is convenient to introduce a superpartner coordinate $\theta$ corresponding to time $t$. 
This permits one to introduce a superfield $Z^I(t, \theta)$ by means of
\begin{eqnarray}
Z^I(t, \theta) = z^I(t) + \theta w^I(t). \label{Z}
\end{eqnarray}
The BV symplectic form just becomes a super extension of 
the BFV symplectic form \eqref{BFVsymplecticFHGD},
\begin{eqnarray}
\omega_{BV} = \int_{T[1]\bR} \!\!\!\! d \theta dt \, \omega_{IJ}(Z) \delta Z^I \wedge \delta Z^J,
\label{BVsymplecticFHGD}
\end{eqnarray}
where $\delta$ denotes the de Rham differential on the extended BV space of fields.

Assume that the BFV symplectic is exact, $\omega_{BFV} = - \delta \vartheta_{BFV}$ for some local $1$-form, \footnote{Formulas in non exact cases appeared in 
\cite{Ikeda:2020eft}. The BV action functional includes the WZ term.} 
\begin{eqnarray} \label{thetaBFV}
\vartheta_{BFV} &=& \vartheta_I(z) \rd z^I,
\end{eqnarray}
parametrized by $\vartheta_I(z)$. 
Then one can define the BV action functional $S_{BV}$ as follows:
\begin{align}
S_{BV}
&:= \int_{T[1]\bR} 
\!\!\!\! d \theta d t \, \vartheta_I(Z) \bbd Z^I
- \int_{T[1]\bR} \!\!\!\! d \theta d t \, (S_{BFV}(Z)+ \theta^0 H_{BFV}(Z)).
\label{FHGDBVaction}
\end{align}
Here $\bbd \equiv \theta \frac{d}{dt}$ can be viewed as the de Rham differential on the line $\bR$ or the corresponding odd and nilpotent vector field on its super extension $T[1]\bR$. 
After integrating out the odd variable $\theta^0$, \eqref{FHGDBVaction} becomes a functional for the fields on $\bR$.
It satisfies the equation, $(S_{BV}, S_{BV}) = 0$
from equations \eqref{BFVequation1}--\eqref{BFVequation3}.
Here $(-,-)$ is the BV bracket induced from the BV symplectic form
\eqref{BVsymplectic1}. 

%%%%%%%%%%%%%%%%%%%%%%%%%%%%%%%%%%%%%%%%%%%%%%%%%%%%%%%%%%%%%%%%%%%%%
%%%%%%%%%%%%%%%%%%%%%%%%%%%%%%   SEC      %%%%%%%%%%%%%%%%%%%%%%%%%%
%%%%%%%%%%%%%%%%%%%%%%%%%%%%%%%%%%%%%%%%%%%%%%%%%%%%%%%%%%%%%%%%%%%%%
\subsection{BV for homogeneous constrained system}\label{BVhomo}
First we consider the homogeneous Hamiltonian system 
in Section \ref{homogeneous} and \ref{free}.

We apply the above procedure to the constrained Hamiltonian mechanics.
The parameter space of time is extended to the super time space $T[1]\bR$ by introducing an odd super time $\theta$. All the coordinates $(x^i, p_i, \eta^a)$ on the phase space of the BFV formalism $T^*[2]E[1]$ is 
extended to super coordinates $(X^i, P_i, Y^a)$, where 
\begin{eqnarray}
X^i(t, \theta) &=& x^i - \theta p^{*i},
\\
P_i(t, \theta) &=& p_i + \theta x^{*}_i,
\\
Y^a(t, \theta) &=& \eta^a - \theta \eta^{*a},
\end{eqnarray}
where $p^{*i}$ and $x^*_i$ are odd coordinates and
$\eta^{*a}$ is an even coordinate.
The BV phase space is the mapping space 
\begin{eqnarray}
\Map(T[1]\bR, T^*[2]E[1]).
\end{eqnarray}
The BV symplectic form is the super extension of the BFV symplectic form
\eqref{BFVsymplectic},
\begin{eqnarray} \label{BVsymplectic}
\omega_{BV} &=& \int_{T[1]\bR} 
\!\!\!\! d\theta dt
\left(\delta X^i \wedge \delta P_i + \frac{1}{2} k_{ab}(X) \delta Y^a \wedge \delta Y^b \right)
\nonumber \\ 
&=& \int_{\bR} 
dt 
\left(\delta x^i \wedge \delta x^*_i + \delta p_i \wedge \delta p^{*i} 
+ k_{ab}(x) \delta \eta^a \wedge \delta \eta^{*b}
- \frac{1}{2} \partial_i k_{ab}(x) p^{*i} \delta \eta^a \wedge \delta \eta^b \right).
\end{eqnarray}
If we assume that $k_{ab}$ is a constant for simplicity,
$\omega_{BFV}$ is an exact form. Then,
the BV action functional $S_{BV}$ is constructed using the formula
\eqref{FHGDBVaction} as follows.
\begin{eqnarray} \label{BVaction}
S_{BV} &=& \int_{T[1]\bR} 
\!\!\!\! d\theta dt [P_i \sd X^i - \frac{1}{2} k_{ab} Y^a  \sd Y^b 
- (S_{BFV}(X, P, Y) + \theta H_{BFV}(X, P, Y))]
\nonumber \\
&=& 
\int_{\bR} 
dt 
\left[p_i \dot{x}^i - \frac{1}{2} k_{ab} \eta^a \dot{\eta}^b 
+ \lambda^a G_a + x^*_i \rho^i_a(x) \eta^a
- p^{*i} \partial_i \rho^j_a(x) p_j \eta^a
- \frac{1}{2} f_{abc}(x) \lambda^a \eta^b \eta^c 
\right.
\nonumber \\
&& 
\left.
- \frac{1}{3!} \partial_i f_{abc}(x) p^{*i} \eta^a \eta^b \eta^c
- \frac{1}{2} g^{ij}(x) p_i p_j
- \frac{1}{2} g^{ij}(x) \Gamma_{abi} p_j \eta^a \eta^b
\right.
\nonumber \\
&& 
\left.
- \frac{1}{8} g^{ij}(x) \Gamma_{abi} \Gamma_{cdj} \eta^a \eta^b \eta^c \eta^d
- \frac{1}{4!} U_{abcd}(x) \eta^a \eta^b \eta^c \eta^d
\right],
\label{BVaction}
\end{eqnarray}
where $\sd = \theta \frac{d}{dt}$ is the superderivative
and $\eta^{*a} = \lambda$.
Here we substitute the concrete BFV functional \eqref{SBFV} 
and the BFV Hamiltonian \eqref{HBFV}.
If all antifields and ghosts are zero, $x^*_i = p^*_i = \eta^a = 0$,
Equation \eqref{BVaction} reduces to the expected classical action \eqref{classicalaction}.

Note that the BV action functional \eqref{BVaction} is not necessarily the same as the BV action functional constructed in the traditional BV procedure from the gauge transformations.
Ambiguity of the FHGD formalism was discussed in \cite{Ikeda:2020eft}.
%The BV action functional \eqref{BVaction} is regarded as the reduction to one dimensional space $\bR$ of the three dimensional Courant sigma model.

An interesting feature is that if we set $x^*_i = p^*_i = \lambda = 0$,
the $S_{BV}$ gives a supersymmetric mechanics
with a Courant algebroid structure.
The 'supersymmetry' is generated by $Q= (S_{BV}, -)$.
Further analysis of the super mechanics leaves in future research.

%%%%%%%%%%%%%%%%%%%%%%%%%%%%%%%%%%%%%%%%%%%%%%%%%%%%%%%%%%%%%%%%%%%%%
%%%%%%%%%%%%%%%%%%%%%%%%%%%%%%   SEC      %%%%%%%%%%%%%%%%%%%%%%%%%%
%%%%%%%%%%%%%%%%%%%%%%%%%%%%%%%%%%%%%%%%%%%%%%%%%%%%%%%%%%%%%%%%%%%%%
\subsection{BV for inhomogeneous constrained system}\label{BVinhomo}
We apply the FHGD method to the BFV formalism 
for the inhomogeneous Hamiltonian mechanics in Section \eqref{BFVinhomo}.

$p_i$ is deformed to $p_i'= p_i + A_i$.
The BFV symplectic form $\omega_{BFV}$, the BFV functional $S_{BFV}$ and the BFV Hamiltonian $H_{BFV}$ are deformed to \eqref{inhomoBFV}, 
\eqref{SBFVinhomo} and \eqref{inhomoHBFV}, respectively.

The parameter superspace is the same as in the homogeneous case.
Super coordinates on the BV phase space $\Map(T[1]\bR, T^*[2]E[1])$ are 
$(X^i, P_i', Y^a)$ which correspond to coordinates 
$(x^i, p_i', \eta^a)$ on the phase space of the BFV formalism,
They are expanded as
\if0
\begin{eqnarray}
X^i(t, \theta) &=& x^i - \theta p^{*i},
\\
P_i'(t, \theta) &=& p_i' + \theta x_i'^{*},
\\
Y^a(t, \theta) &=& \eta^a - \theta \eta^{*a}.
\end{eqnarray}
\fi
The BV symplectic form is the super extension of the BFV symplectic form 
\eqref{BFVsymplectic},
\begin{eqnarray} \label{BVsymplectic}
\omega_{BV} &=& \int_{T[1]\bR} 
\!\!\!\! d\theta dt
\left(\delta X^i \wedge \delta P_i'
+ \frac{1}{2} B_{ij}(X) \delta X^i \wedge \delta X^j
+ \frac{1}{2} k_{ab}(X) \delta Y^a \wedge \delta Y^b \right)
\nonumber \\ 
&=& \int_{\bR} 
dt 
\left(\delta x^i \wedge \delta x'{}^*_i + \delta p_i' \wedge \delta p^{*i} 
+ B_{ij}(x) \delta x^i \wedge \delta p^{*j}
- \frac{1}{2} \partial_k B_{ij}(x) p^{*k} \delta x^i \wedge \delta x^j
\right.
\nonumber \\ &&
\left.
+ k_{ab}(x) \delta \eta^a \wedge \delta \eta^{*b}
- \frac{1}{2} \partial_i k_{ab}(x) p^{*i} \delta \eta^a \wedge \delta \eta^b \right).
\end{eqnarray}
For the symplectic form $\omega_{BFV}$, the Liouville $1$-form $\vartheta$ such that $\omega_{BV} = - d \vartheta$ is locally written as
\begin{eqnarray} \label{BFVLiouville}
\vartheta_{BFV} &=& p_i' \rd x^i - A_i(x) \rd x^i - \frac{1}{2} k_{ab} \eta^a \rd \eta^b.
\end{eqnarray}
Finally, the BV action functional $S_{BV}$ is obtained from
the formula in the FHGD formulation
by substituting concrete expressions of the BFV functional \eqref{SBFV} 
and the BFV Hamiltonian \eqref{HBFV},
\begin{eqnarray} \label{inhomoBVaction}
S_{BV} &=& \int_{T[1]\bR} 
\!\!\!\! d\theta dt 
[P_i' \sd X^i - A_i(X) \sd X^i - \frac{1}{2} k_{ab} Y^a \sd Y^b 
- (S_{BFV}(X, P', Y) + \theta H_{BFV}(X, P', Y))]
\nonumber \\
&=& 
\int_{\bR} dt 
\left[p_i' \dot{x} - \frac{1}{2} k_{ab} \eta^a \dot{\eta}^b\right]
+ \int_{T[1]\Sigma} d\theta dtd\sigma B(x)
\nonumber \\ && 
+ \int_{T[1]\bR} d\theta dt 
\left[\lambda^a (\rho^i_a p_i' + \mu_a) + x'{}^*_i \rho^i_a(x) \eta^a
- p^{*i} (\partial_i \rho^j_a(x) p_j' + \partial_i \mu_a(x)) \eta^a
\right.
\nonumber \\
&& 
\left.
- \frac{1}{2} f_{abc}(x) \lambda^a \eta^b \eta^c 
- \frac{1}{3!} \partial_i f_{abc}(x) p^{*i} \eta^a \eta^b \eta^c
- \frac{1}{2} g^{ij}(x) p_i' p_j'
- V'(x)
\right.
\nonumber \\
&& 
\left.
- \frac{1}{2} g^{ij}(x) \Gamma_{abi} p_j' \eta^a \eta^b
- \frac{1}{8} g^{ij}(x) \Gamma_{abi} \Gamma_{cdj} \eta^a \eta^b \eta^c \eta^d
- \frac{1}{4!} U_{abcd}(x) \eta^a \eta^b \eta^c \eta^d
\right],
\label{BVaction}
\end{eqnarray}
where $\Sigma$ is a two dimensional manifold such that $\bR = \partial \Sigma$
with a local coordinate $(t, \sigma)$.
The second term in \eqref{BFVLiouville} gives the WZ term on $T[1]\Sigma$ since
$B = dA$.
If all antifields and ghosts are zero, $x^*_i = p^*_i = \eta^a = 0$,
Equation \eqref{BVaction} reduces to the classical action \eqref{classicalaction}.

%%%%%%%%%%%%%%%%%%%%%%%%%%%%%%%%%%%%%%%%%%%%%%%%%%%%%%%%%%%%%%%%%%%%%
%%%%%%%%%%%%%%%%%%%%%%%%%%%%%%   SEC      %%%%%%%%%%%%%%%%%%%%%%%%%%
%%%%%%%%%%%%%%%%%%%%%%%%%%%%%%%%%%%%%%%%%%%%%%%%%%%%%%%%%%%%%%%%%%%%%
\section{Weil algebra}\label{Weilalgebra}
The Weil algebra is a model for the equivariant cohomology
based on a graded algebra 
We refer to Kalkman's formulation \cite{Kalkman:1993zp, Kalkman93} and the textbook \cite{Guillemin:1999ge}.

We construct a Weil algebra from a Q-manifold.
A Weil algebra on a Q-manifold has been discussed in \cite{Mehta09, AbadCrainic11}.
We need only the homological vector field $Q$, however, 
we consider a QP-manifold by considering a graded symplectic structure. 
We can easily connect the Weil algebra with the BFV and BV formalism
since the both BFV and BV formalism have QP-manifolds.
%it is useful to also assume a graded symplectic form $\omega$ 
%since the BFV and BV formalisms have symplectic forms and homological 
%functions which is a Hamiltonian of homological vector fields.
%An underlining mathematical structure for the BFV and BV is a QP-manifold.
%We consider derivation of the Weil algebra from the BFV or BV structures.

%%%%%%%%%%%%%%%%%%%%%%%%%%%%%%%%%%%%%%%%%%%%%%%%%%%%%%%%%%%%%%%%%%%%%
%%%%%%%%%%%%%%%%%%%%%%%%%%%%%%   SEC      %%%%%%%%%%%%%%%%%%%%%%%%%%
%%%%%%%%%%%%%%%%%%%%%%%%%%%%%%%%%%%%%%%%%%%%%%%%%%%%%%%%%%%%%%%%%%%%%
\subsection{Weil model on Q-manifold}

%%%%%%%%%%%%%%%%%%%%%%%%%%%%%%%%%%%%%%%%%%%%%%%%%%%%%%%%%%%%%%%%%%%%%
%%%%%%%%%%%%%%%%%%%%%%%%%%%%%%   SEC      %%%%%%%%%%%%%%%%%%%%%%%%%%
%%%%%%%%%%%%%%%%%%%%%%%%%%%%%%%%%%%%%%%%%%%%%%%%%%%%%%%%%%%%%%%%%%%%%
%\subsection{QP-manifold}
%\noindent
\if0
A graded manifold $\calM$ is a ringed space with a structure sheaf
of 
%a nonnegatively 
$\bZ$-graded commutative
algebra over an ordinary smooth manifold $M$.
Grading is compatible with supermanifold grading,
that is, a variable of even degree is commutative, and
one of odd degree is anticommutative. 
By definition, the structure sheaf of $\calM$ is locally isomorphic to
$C^{\infty}(U) \otimes S^{\bullet}(V)$,
where
$U$ is a local chart on $M$,
$V$ is a graded vector space, and $S^{\bullet}(V)$ is a free
graded commutative ring on $V$.
%
Refer to \cite{Carmeli}\cite{Varadarajan}
for the rigorous mathematical definition of objects
in supergeometry.
We use conventions in \cite{Ikeda:2012pv}.
\fi

A graded manifold $\calM$ is called an \textit{N-manifold} if 
a graded manifold is a nonnegatively graded.
\if0
An \textit{N-manifold} 
equipped with a graded symplectic structure
$\omega$ of degree $n$ is
%denoted by $(\calM, \omega)$.
A graded Poisson bracket on $C^\infty ({\calM})$ is defined as
\beq
\{f,g\} = (-1)^{|f|+n+1} i_{X_f} i_{X_g}\omega
\eeq
where a Hamiltonian vector field $X_f$ is defined by the equation
$i_{X_f}\omega= - \delta f$
for any $f\in C^\infty({\calM})$, and
$\delta$ is a differential on $\calM$.
A vector field $Q$ on $\calM$ is called the homological vector field 
if $Q^2=0$.
\fi
\begin{definition}
A \textit{QP-manifold (differential graded symplectic manifold)
} 
is a $N$-manifold $\calM$ with a graded symplectic form $\omega$
endowed with a degree $1$ homological vector field $Q$ 
such that $L_Q \omega =0$, 
where $L_Q$ is a graded Lie derivative.
\cite{Schwarz:1992nx}.
\end{definition}
%We also denote a QP-manifold by
%the corresponding triple $(\calM,\omega,Q)$.
A graded symplectic form gives a graded Poisson bracket $\{-,-\}$.
For any QP manifold of positive degree, there exists 
a Hamiltonian function $\Theta\in C^{\infty}(\calM)$ of $Q$ 
with respect to the graded Poisson bracket $\{-,-\}$,
\beq
Q=\{\Theta,-\}.
\eeq
Then, the homological condition of $Q$, $Q^2=0$, implies that
$\Theta$ is a solution of the equation
\begin{equation}
\{\Theta,\Theta\}=0.
\label{cmaseq}
\end{equation}
%Such $\Theta$ is also called a homological function.
%A Q-manifold $(\calM, Q)$
%corresponds to BRST models in the equivariant cohomology theory.

%%%%%%%%%%%%%%%%%%%%%%%%%%%%%%%%%%%%%%%%%%%%%%%%%%%%%%%%%%%%%%%%%%%%%
%%%%%%%%%%%%%%%%%%%%%%%%%%%%%%   SEC      %%%%%%%%%%%%%%%%%%%%%%%%%%
%%%%%%%%%%%%%%%%%%%%%%%%%%%%%%%%%%%%%%%%%%%%%%%%%%%%%%%%%%%%%%%%%%%%%
%\subsection{Weil model on Q-manifold}

A Weil algebra $(W, \rd, \iota, L)$ is a graded algebra $W$ with three derivations $\rd, \iota, L$.
We construct a Weil algebra from a QP-manifold $\calM$.
The space of a Weil algebra is $W = C^{\infty}(T[1]\calM)$.
Three operations $(\rd, \iota, L)$ are 
%defined as derivations on $W$,where they 
are derivations of degree $(1, -1, 0)$ on $W$.

Take a local coordinate $(e^a, \theta^a)$ on $T[1]\calM$
of degree $|\theta| = |e|+1$.
%For a differential of function $e \in C^{\infty}(\calM)$, 
%$\delta e = \delta e^a \frac{\delta e}{\delta e^a}$,
%where $\delta$ is the exterior derivative on $\calM$.
We denote the superderivative of an element $e \in C^{\infty}(\calM)$
by $\Fse := \delta e = \theta^a \frac{\partial e}{\partial e^a}$
called a tangent vector along $e$.
Note that $\Fse$ is a linear function of the tangent direction
of $T[1]\calM$.
%Especially, we denote the basis corresponding to $e^a$ by $\Fse{}^a = \theta^a$.

For functions $e, e_1, e_2 \in C^{\infty}(\calM)$ on a base graded manifold,
we define $\rd, \iota_e, L_e: C^{\infty}(\calM) \rightarrow C^{\infty}(\calM)$
%using the graded Poisson bracket $\{-,-\}$ and the homological function $\Theta$ on the QP-manifold $\calM$ 
as follows,
\begin{eqnarray}
&& \rd e := \Fse + Q{e} = \delta e + Q{e}.
\label{Weilalgebra1}
\\
&& \iota_{e_1}(e_2) := \sbv{e_1}{e_2},
\label{Weilalgebra2}
\\
&& 
L_{e_1}(e_2) := \sbv{\sbv{e_1}{\Theta}}{e_2},
%L_{e_1}(e_2) = \sbv{e_1}{\sbv{\Theta}{e_2}},
\label{Weilalgebra3}
\end{eqnarray}
From the definition, $d, \iota, L$ is of degree $1, -1, 0$, respectively.
%From Equation (\ref{Weilalgebra1}), the Weil differential 
%$\rd$ is regarded as a connection on $\calM$.
%$Q$ corresponds to the BRST differential in the BRST model.
For a basis $e^a$ on $\calM$, Equation (\ref{Weilalgebra1}) is 
%\begin{eqnarray}
$\rd e^a = \theta^a + Q{e^a}.$
%\label{Weilalgebra33}
%\end{eqnarray}
$e$ is regraded as a 'connection' and $\Fse = \rd e - Q{e}$
is a 'curvature'.

Three operations $\rd, \iota, L$ on the tangent direction
are defined as follows.
%the basis $\theta^a$ in 
We require that the 'differential' $\rd$ satisfies $\rd^2=0$,
and the 'Lie derivative' $L_e$ satisfies the Cartan magic formula, 
$L_e = \iota_e \rd - (-1)^{|\iota_e|} \rd \iota_e
= \iota_e \rd + (-1)^{|e|} \rd \iota_e$, respectively,
where $|\iota_e| = |e|-1$ is degree of $\iota_e$.
From Equation \eqref{Weilalgebra3} on the basis 
and $\rd^2=0$, we obtain Equation,
\begin{eqnarray}
\rd \theta^a &=& - \rd Qe^a = - \delta Qe^a.
\label{Weilalgebra4}
\end{eqnarray}
since $Q^2=0$ and $\Fse = (\rd - Q)e = \delta e$.
From the requirement $L_e = \iota_e \rd + (-1)^{|e|} \rd \iota_e$,
the following definitions of $\iota \theta^a$ and $L \theta^a$ are obtained,
\begin{eqnarray}
\iota_{e_1} \theta^a
&=& - (-1)^{|{e_1}|} (\rd - Q)\iota_{e_1} e^a
= -(-1)^{|{e_1}|} \delta \iota_{e_1} e^a,
%&=& (-1)^{|\iota_{e_1}|} (\rd - Q)\iota_{e_1} e^a
%= (-1)^{|\iota_{e_1}|} \delta \iota_{e_1} e^a,
\label{Weilalgebra5}
\\
L_{e_1} \theta^a 
&=& (-1)^{|{e_1}|} (\rd - Q) L_{e_1} e^a
= (-1)^{|{e_1}|} \delta L_{e_1} e^a.
%&=& - (-1)^{|\iota_{e_1}|} (\rd - Q) L_{e_1} e^a
%= - (-1)^{|\iota_{e_1}|} \delta L_{e_1} e^a.
\label{Weilalgebra6}
\end{eqnarray}
%
%By requiring derivation properties,
%i.e. the Leibniz rule for each operation, 
Equations \eqref{Weilalgebra1}--\eqref{Weilalgebra3} and 
\eqref{Weilalgebra4}--\eqref{Weilalgebra6} fix 
all operations of $(\rd, \iota, L)$ on elements $W = C^{\infty}(T[1]\calM)$,
thus we obtain the Weil algebra from a QP-manifold.

We summarize useful formulas calculated the above definitions of operations $(\rd, \iota, L)$.
For the element $\Fse = \delta e$, the following formulas
are obtained.
\begin{eqnarray}
\rd \Fse = - \rd Q{e} = - \Fsqe,
\label{Weilalgebra7}
\end{eqnarray}
since $Q^2=0$.
From the Cartan magic formula, we obtain
\begin{eqnarray}
\iota_{e_1} F_{e_2}
%&=& - (-1)^{|{e_1}|} (\rd - Q)\iota_{e_1} e_2
%\nonumber \\
%&=& (-1)^{|\iota_{e_1}|} F_{\iota_{e_1}{e_2}},
%\nonumber \\
&=& -(-1)^{|e_1|} F_{\iota_{e_1}{e_2}},
\label{Weilalgebra8}
\\
L_{e_1} F_{e_2} 
%&=& - \iota_{e_1} \rd Q {e_2} + \rd F_{\sbv{e_1}{e_2}}
%\nonumber \\
%&=& 
%(-1)^{|{e_1}|} (\rd - Q) L_{e_1}{e_2}
%\nonumber \\
%&=& 
%(-1)^{|L_{e_1}|} F_{L_{e_1}{e_2}}
%\nonumber \\
&=& 
- (-1)^{|{e_1}|} F_{L_{e_1}{e_2}}.
%- (-1)^{|\iota_{e_1}|} F_{\iota_{e_1}Q e_2} - F_{Q \iota_{e_1} e_2}.
\label{Weilalgebra9}
\end{eqnarray}
Operations $(\rd, \iota, L)$ consist of a graded Lie algebra 
with a graded Lie bracket $[-,-]$ as follows,
\begin{eqnarray}
~[\iota_{e_1}, \iota_{e_2}] &=& \iota_{\iota_{e_1}{e_2}},
\label{gradedLiebracket1}
\\
~[\iota_{e_1}, L_{e_2}] 
&=& \iota_{(L_{e_1}{e_2})},
\label{gradedLiebracket2}
\\
~[L_{e_1}, L_{e_2}] 
&=& L_{(L_{e_1}{e_2})}.
\label{gradedLiebracket3}
\end{eqnarray}
%We can concretely prove  Equations (\ref{gradedLiebracket1})--(\ref{gradedLiebracket3}) by operating each action to elements of $W$.
%We call consist of a Weil algebra on $C^{\infty}(T[1]\calM)$.
%\begin{theorem}
\if0
An element $- (-1)^{|e|-n}Qe = \sbv{e}{\Theta}$ is the moment map with respect to $\omega$
because it is a Hamiltonian function for the action $L_{e}$ 
and consist of a (graded) Lie algebra from Equation
(\ref{gradedLiebracket3}).
\fi
$\courant{e_1}{e_2} := L_{e_1}{e_2} = \sbv{\sbv{e_1}{\Theta}}{e_2}$
gives a bilinear bracket.
In fact, since this bracket satisfies the Leibniz identity 
\eqref{Leibnizidentity},
but not necessarily skew symmetric,
%$\courant{-}{-}$
%\cite{Dorfman}
it is the Dorfman bracket \cite{Kosmann-Schwarzbach:2003en}.
Thus we can also write (\ref{gradedLiebracket2}) and (\ref{gradedLiebracket3})
as
\begin{eqnarray}
~[\iota_{e_1}, L_{e_2}] 
&=& \iota_{\courant{e_1}{e_2}},
\\
~[L_{e_1}, L_{e_2}] 
&=& L_{\courant{e_1}{e_2}}.
\end{eqnarray}
%$(L_e, \iota_e, d)$ satisfies the Cartan formula $d^2=0$ and $L_e = \iota_e d - (-1)^{|\iota_e|} d \iota_e$. 

\begin{definition}
We call the graded Lie algebra acting on the space $W= C^{\infty}(T[1]\calM)$
generated by $(\rd, \iota_e, L_e)$ defined by Equations (\ref{Weilalgebra1}) -- (\ref{Weilalgebra6}) 
a Weil algebra for a graded (QP-)manifold $\calM$.
\end{definition}
%\end{theorem}

Since $\omega$ is symplectic, $\iota_e = \sbv{e}{-}$ is nondegenerate, 
and thus the Weil algebra satisfies the property corresponding to a locally
free action called \textit{type (C)} in \cite{Guillemin:1999ge}.

\if0
We see that our Weil algebra satisfies
the following property corresponding to a locally free action
\begin{definition}
A a graded algebra or a module of a graded algebra
is called \textbf{type (C)}
if operations $\iota_{e}$ is nondegenerate.
\end{definition}
\begin{theorem}
Since $\omega$ is symplectic,
$\iota_e = \sbv{e}{-}$ is nondegenerate.
%Therefore the above Weil algebra $W$ is of type (C).
\end{theorem}
\fi
\begin{definition}
Let $\varphi \in W$.
\begin{enumerate}
\item If $\iota_e \varphi =0$ for any $e \in C^{\infty}(\calM)$,
$\varphi$ is called \textit{horizontal}.
\qquad $W_{hor} = \{\gamma \in W | \iota_e \varphi =0 \}$.
\item 
If $\varphi$ satisfies $L_e \gamma = \iota_e \varphi =0$
for any $e \in C^{\infty}(\calM)$, 
it is called \textit{basic}. 
$W_{bas} = \{ \varphi \in W| L_e \varphi = \iota_e \varphi =0 \}$.
%Then $\gamma$ is an element of the basic cohomology ring 
%$W_{bas} = W_{hor}^{\calG}$.
\end{enumerate}
\end{definition}

%%%%%%%%%%%%%%%%%%%%%%%%%%%%%%%%%%%%%%%%%%%%%%%%%%%%%%%%%%%%%%%%%%%%
%%%%%%%%%%%%%%%%%%%%%%%%%%%%%%%%%%%%%%%%%%%%%%%%%%%%%%%%%%%%%%%%%%%%
%%%%%%%%%%%%%%%%%%%%%%%%%%%%%%%%%%%%%%%%%%%%%%%%%%%%%%%%%%%%%%%%%%%%
\subsection{Cartan Model}\label{cartanmodel}
\noindent
Let $B$ be a (graded) $\calM$-module.
$\calM$ acts on $B$ as infinitesimal actions, i.e., 
operations $(\rd, \iota, L)$.
%which is a module with an infinitesimal action of the QP-manfold $\calM$ to $B$.

The space of the Cartan model is $W \otimes B$.
Three operations on $W \otimes B$ are given by
\begin{eqnarray}
&&
L_{e} = L_{e} \otimes 1 + 1 \otimes L_{e},
\\
&& 
\iota_e = \iota_e \otimes 1 + 1 \otimes \iota_e,
\\
&& 
d = d \otimes 1 + 1 \otimes d.
\end{eqnarray}
$(W \otimes B)_{hor} = \{ \varphi \in W \otimes B | \iota \varphi = 0\}$
is the space of horizontal functions and 
$(W \otimes B)_{bas} = \{ \varphi \in W \otimes B 
| \iota \varphi = L \varphi = 0 \}$
is the space of basic functions.

We can construct the generator of the Mathai-Quillen map from 
the space of the Weil model to the space of the Cartan model
\cite{Guillemin:1999ge}.
It is defined by 
\begin{eqnarray}
\phi 
%= \exp \iota_Q 
= \exp\left(h \otimes \iota
\right).
\end{eqnarray}
%\cite{Mehta09}
%$\phi = \exp (p_{a(i)} \otimes \iota^{a(i)}
%- (-1)^{i(n-i)} q^{a(i)} \otimes \iota_{a(i)})$.
\if0
Now we consider the adjoint action:
$\phi X \phi^{-1} 
= \sum_{m=0}^{\infty} \frac{1}{m!} 
{\rm ad}^m
\left(h \otimes \iota
\right)
X$ 
for operators $X$ on $W \otimes B$.
\fi
Here $h$ is an operator such that 
%\begin{eqnarray}
%&& 
$[h, \iota] \varphi = - \varphi$,
%\end{eqnarray}
for a horizontal function $\varphi \in W$.
The computation gives
$\rd \varphi - h L \varphi = 0.$
%\end{eqnarray}
%
%
\if0
%$f(\Fsq^{a(i)}, \Fsp_{a(i)})$.
%From equations (\ref{iotaLcom1})-(\ref{iotaLcom4}),
Equation (\ref{ad2}) is equal to 
$1 \otimes Q$ for functions on the base manifold
and $0$ for the tangent elements on $T[1]\calM$.
Using these formulas and $\rd  - Q =0$
on the horizontal subspace,
%simply Equations (\ref{ad1})-(\ref{ad3}).
\fi
%
%
Thus, we obtain the equivariant differential 
$\rd_C$ such that $\rd_C^2=0$ for the Cartan model as
\begin{eqnarray}
\rd_C &=& 
\phi (\rd \otimes 1 + 1 \otimes \rd) \phi^{-1}  =
1 \otimes \rd 
- (\rd - Q)e^a \otimes \iota_{e^a}
= 1 \otimes \rd - \theta^a \otimes \iota_{e^a}.
%= 1 \otimes \rd - F_{e^a} \otimes \iota_{e^a}.
%- \Fsp_{a(i)} \otimes \iota^{a(i)}
%+ (-1)^{i(n-i)} \Fsq^{a(i)} \otimes \iota_{a(i)}.
%\sim Q.
\end{eqnarray}

We have the following 
equivalence of the Weil model and the Cartan model.
Let $S$ be the horizontal subspace 
of $W$
%spanned by $(\Fsq, \Fsp)$, 
and $C_{\calG}(B) = (S \times B)^{\calG}$
is an invariant subspace of the action of $\calG$
whose infinitesimal action is defined by $\calM$.
\footnote{For $\calG$, the integration of the graded manifold $T^*[2]E[1]$ and a Courant algebroid to a group-like object has been analyzed in \cite{LBS:2011, MT:2013}.
A Lie rackoid is also proposed as an integration of a Courant algebroid
\cite{Laurent-Gengoux:2015, Laurent-Gengoux:2018zoz}.}
%\begin{theorem}
We consider the basic subspace,
%\begin{eqnarray}
$(W \otimes B)_{bas}$.
%=\{ \varphi \in W \otimes B | L{\varphi} = \iota \varphi =0 \}.
%\end{eqnarray}
On this space the Mathai-Quillen isomorphism shows equivalence of 
the cohomologies of the Weil model and the Cartan model
$\phi$ maps
$(W \otimes B)_{bas}$ into $C_{\calG}(B)$ and $\rd$ to $\rd_C$.  
Thus we obtain following equivalence of 
equivariant cohomology computed from 
the Weil model and the Cartan model.
\begin{eqnarray}
H^*((W \otimes B)_{bas},\rd) = H^*(C_{\calG}(B), \rd_C).
\end{eqnarray}
%\end{theorem}

%%%%%%%%%%%%%%%%%%%%%%%%%%%%%%%%%%%%%%%%%%%%%%%%%%%%%%%%%%%%%%%%%%%%%
%%%%%%%%%%%%%%%%%%%%%%%%%%%%%%   SEC      %%%%%%%%%%%%%%%%%%%%%%%%%%
%%%%%%%%%%%%%%%%%%%%%%%%%%%%%%%%%%%%%%%%%%%%%%%%%%%%%%%%%%%%%%%%%%%%%
\subsection{Weil model and Cartan model for Courant algebroid}\label{weilcartanhomo}
In this section, we compute concrete formulas.
For the Courant algebroid, we consider the QP-manifold for the Courant algebra 
$\calM= T^*[2]E[1]$ with the homological vector field $Q$.
The homological function $\Theta$ is $S_{BFV}$ in Equation \eqref{SBFVhomo}.

The Weil model is defined on the graded algebra $W = C^{\infty}(T[1]T^*[2]E[1])$.

For local coordinates $(x^i, \eta^a, p_i)$ on $T^*[2]E[1]$,
ones on the tangent direction $T[1]$ are $(\Fsx^i, \Fsq^a, \Fsxi_i)$
of degree $(1, 2, 3)$.

Substituting Equation \eqref{SBFVhomo} to the definitions \eqref{Weilalgebra1}--\eqref{Weilalgebra3}, we obtain the Weil differential $\rd$ is
\begin{eqnarray}
&&
\rd x^i = \Fsx^i - \rho^{i}{}_a \eta^a,
\label{CAcurvature1}
\\
&& 
\rd \eta^a = \Fsq^a 
+ k^{ab} \rho^{i}{}_b p_i^{\nabla}
+ k^{ab} \Gamma_{bci} \rho^{i}{}_d \eta^c \eta^d
- \frac{1}{2} k^{ab} T_{bcd} \eta^c \eta^d,
%+ k^{ab} \rho^{i}{}_b p_i
%- \frac{1}{2} k^{ab} f_{bcd} \eta^c \eta^d,
\label{CAcurvature2}
\\
&& 
\rd p_i^{\nabla} = \Fsxi^{\nabla} _i
+ D_i \rho^{j}{}_a \eta^a p_j^{\nabla}
- \frac{1}{3!} S_{iab}^d k_{cd} \eta^a \eta^b \eta^c.
%+ \partial_i \rho^{j}{}_a p_j \eta^a
%- \frac{1}{3!} \partial_i f_{abc} \eta^a \eta^b \eta^c.
\label{CAcurvature3}
\end{eqnarray}
Note that the Poisson bracket of the basis $e_a$ on $E$ with $p_i^{\nabla}$ is not zero as \eqref{Poissonpe}, $e_a$ has a nontrivial transformation under $Q$.
$\eta = \eta^a e_a$ covariantly transforms under operations of 
the Weil algebra. 
For $\Fsx = \Fsx^i \partial_i, \Fsq = \Fsq^a e_a, \Fsxi = \Fsxi_i d x^i$, we obtain covariant formulas,
\begin{eqnarray}
&&
\rd x = \Fsx - \rho(\eta),
\label{CAcurvature1b}
\\
&& 
\rd \eta = \Fsq
+ \iota_{\rho^*} p^{\nabla}
- T^*(\eta, \eta),
%+ k^{ab} \rho^{i}{}_b p_i
%- \frac{1}{2} k^{ab} f_{bcd} \eta^c \eta^d,
\label{CAcurvature2b}
\\
&& 
\rd p^{\nabla} = \Fsxi^{\nabla}
+ p^{\nabla}(D \rho(\eta))
- S^*(\eta, \eta, \eta),
%+ \partial_i \rho^{j}{}_a p_j \eta^a
%- \frac{1}{3!} \partial_i f_{abc} \eta^a \eta^b \eta^c.
\label{CAcurvature3b}
\end{eqnarray}
where $x = x^i \partial_i, \eta = \eta^a e_a, p^{\nabla} = p_i^{\nabla} dx^i$.
$\rho^* \in \Gamma(TM \oplus E)$, $T^* \in \Gamma(\wedge^2 E^* \oplus E)$ 
and $S^* \in \Gamma(T^*M \oplus \wedge^3 E^*)$ are defined by
$\bracket{\rho^*}{e} = \rho(e)$,
$\bracket{T^*}{e}(-,-) = T(-,-,e)$ and
$S^*(-,-,e) = \bracket{S}{-,-,e}$ for $e \in \Gamma(E)$.

For the tangent direction, we obtain the Weil differential by
acting $\rd$ to Equations \eqref{CAcurvature1}--\eqref{CAcurvature3},
\begin{eqnarray}
\rd \Fsx &=& \Fsx \rho(\eta) + \rho(\Fsq).
%\partial_j \rho^i{}_a \Fsx^j \eta^a + \rho^i{}_a \Fsq^a,
\\
\rd \Fsq^a &=&
\Fsx_{D} \left[- \iota_{\rho^*} p^{\nabla} + T^*(\eta, \eta) \right] 
- \iota_{\rho^*} \Fsxi + T^*(\Fsq, \eta),
%- \left[\partial_j( k^{ab} \rho^{i}{}_b) p_i
%- \frac{1}{2} \partial_j(k^{ab} f_{bcd}) \eta^c \eta^d \right] \Fsx^j
%- k^{ab} \rho^{i}{}_b \Fsxi_i
%+ k^{ab} f_{bcd} \Fsq^c \eta^d,
\\
\rd \Fsxi^{\nabla} &=&
\iota_{\Fsx} \left[- p^{\nabla} (D \rho^*(\eta)) 
+ S^*(\eta, \eta, \eta)\right]
\nonumber \\
&& 
+ \iota_{D\rho(\eta)} \Fsxi^{\nabla}
+ \left[- p^{\nabla} (D \rho(\Fsq)) + S^*(\Fsq, \eta, \eta) \right].
%&=& \left[-\partial_i \partial_j \rho^k{}_a p_k \eta^a
%+ \frac{1}{3!} \partial_i \partial_j f_{abc} \eta^a \eta^b \eta^c
%\right] \Fsx^j 
%\nonumber \\ && 
%- \partial_i \rho^{j}{}_a \Fsxi_j \eta^a
%- \left[\partial_i \rho^{j}{}_a p_j
%+ \frac{1}{2} f_{abc} \eta^b \eta^c \right] \Fsq^a.
\end{eqnarray}
$\Fsx_{D}$ is a covariantized vector field given by $\Fsx_{D} = \Fsx^i D_i$.
Important formulas from the general theory are $\Fsx^i$, $\Fsq^a$ and $\Fsxi_i$ are horizontal,
\begin{eqnarray}
\iota_e \Fsx^i =\iota_e \Fsq^a = \iota_e \Fsxi_i =0.
\end{eqnarray}

For the Cartan model, we consider the basic subspace,
\begin{eqnarray}
(W \otimes B)_{bas}
=\{ \gamma \in W \otimes B | L{\varphi} = \iota \varphi =0 \}.
\end{eqnarray}
The equivariant differential on the Cartan model on the basic subspace
is 
\begin{eqnarray}
\rd_C 
&=& 1 \otimes d 
- \Fsxi_i \otimes \iota_{p}^i
+ \Fsx^i \otimes \iota_{x i}
- \frac{1}{2} k_{ab} \Fsq^a \otimes \iota_{\eta}^b.
\end{eqnarray}

%%%%%%%%%%%%%%%%%%%%%%%%%%%%%%%%%%%%%%%%%%%%%%%%%%%%%%%%%%%%%%%%%%%%%
%%%%%%%%%%%%%%%%%%%%%%%%%%%%%%   SEC      %%%%%%%%%%%%%%%%%%%%%%%%%%
%%%%%%%%%%%%%%%%%%%%%%%%%%%%%%%%%%%%%%%%%%%%%%%%%%%%%%%%%%%%%%%%%%%%%
\subsection{Weil model and Cartan model with momentum section}\label{weilcartaninhomo}

We choose the inhomogeneous BFV functional $S_{BFV}$ 
in Equation \eqref{SBFVinhomo} as a homological funciton $\Theta$.
The Weil model in Subsection \ref{weilcartanhomo} is deformed by 
the momentum section term.
The Weil differential $\rd$ is deformed to
\begin{eqnarray}
&&
\rd' x^i = \rd x^i,
\label{CAcurvature21}
\\
&&
\rd' \eta^a = \rd \eta^a + k^{ab} \mu_b,
\label{CAcurvature22}
\\
&& 
\rd' p'{}_i^{\nabla} = \rd p'{}_i^{\nabla} 
+ (D_i \mu_a - \gamma_a) \eta^a.
\label{CAcurvature23}
\end{eqnarray}
The coordinate independent form is
\begin{eqnarray}
&&
\rd' x = \rd x ,
\label{CAcurvature1c}
\\
&& 
\rd' \eta = \rd \eta + \mu^*,
\label{CAcurvature2c}
\\
&& 
\rd' p^{\nabla} = \rd p^{\nabla} 
+ D \mu (\eta) - \gamma(\eta),
\label{CAcurvature3c}
\end{eqnarray}
where $\mu^* \in \Gamma(E)$ is defined by $\mu(e) = \bracket{\mu^*}{e}$.
By the straight forward calculation, we obtain the deformations of 
the Weil differentials for the tangent direction,
\begin{eqnarray}
\rd' \Fsx &=& \rd \Fsx + \rho(\mu^*),
%\rd \Fsx^i  + \rho^i_a k^{ab} \mu_b,
\\
\rd' \Fsq &=& \rd \Fsq 
- \iota_{\rho^*}(D \mu - \gamma) (\eta) + T^*(\mu^*, \eta),
\\
\rd' \Fsxi_i &=& \rd \Fsxi_i 
+ \iota_{D \rho(\eta)} (D \mu - \gamma)(\eta)
- \iota_{D\rho(\mu^*)} p^{\nabla} + S^*(\mu^*, \eta, \eta).
\end{eqnarray}
The Weil algebra has been constructed from a QP-manifold.
If $\mu$ is a momentum section, $Q^2=0$ is satisfied.
Thus we obtain the general formula of the Weil algebra.
\begin{theorem}
If $\mu$ is a momentum section,
$\Fsx^i$, $\Fsq^a$ and $\Fsxi_i$ are horizontal,
\begin{eqnarray}
\iota_e \Fsx^i=\iota_e \Fsq^a = \iota_e \Fsxi_i =0.
\end{eqnarray}
\end{theorem}
The equivariant differential on the Cartan model on the basic subspace
is given by
\begin{eqnarray}
\rd_C 
&=& 1 \otimes d 
- \Fsxi_i \otimes \iota_{p}^i
+ \Fsx^i \otimes \iota_{x i}
- \frac{1}{2} k_{ab} \Fsq^a \otimes \iota_{\eta b}.
\end{eqnarray}

%%%%%%%%%%%%%%%%%%%%%%%%%%%%%%%%%%%%%%%%%%%%%%%%%%%%%%%%%%%%%%%%%%%%%
%%%%%%%%%%%%%%%%%%%%%%%%%%%%%%   SEC      %%%%%%%%%%%%%%%%%%%%%%%%%%
%%%%%%%%%%%%%%%%%%%%%%%%%%%%%%%%%%%%%%%%%%%%%%%%%%%%%%%%%%%%%%%%%%%%%
\section{Conclusion and discussion}
We have defined a momentum section on a pre-symplectic manifold with a Courant algebroid and a Hamiltonian Courant algebroid.
They are a generalization of the momentum map theory on a pre-symplectic manifold with a Lie algebra action.

We analyzed the constrained Hamiltonian mechanics as a concrete nontrivial example. We checked that conditions of the momentum section are equivalent to the consistency conditions of the Hamiltonian mechanics.
The BFV and BV formalisms of this mechanics are constructed and a momentum section is described as terms of inhomogeneous degree of the BFV function, the BFV Hamiltonian and the BV functional.
The BFV and BV formalisms give cohomological formulations of momentum sections.
As an application, we consider the Weil algebra to realize this structure.

An important application of the momentum map theory is 
the symplectic reduction \cite{Marsden-Weinstein}. 
Several proposal of 'Groupoid' objects for the Courant algebroid has been in 
\cite{LBS:2011, MT:2013, Laurent-Gengoux:2015, Laurent-Gengoux:2018zoz}.
Since we have constructed a momentum section theory for a Courant algebroid, 
if a 'groupoid' object $\calG$ for the Courant algebroid acts on a symplectic manifold $M$, the quotient space $M/\calG$ should be a symplectic manifold.
We need more analysis since the 'groupoid' object is more complicated than a momentum map case for a Lie group.

Another important but related application is the equivariant cohomology.
The theory is strongly connected to quantizations of physical theories.
If we can obtain a localization formula similar to the Duistermaat-Heckman formula \cite{Duistermaat-Heckman} for our setting, we will be able to apply it to quantum theories with Courant algebroid structures.

%%%%%%%%%%%%%%%%%%%%%%%%%%%%%%%%%%%%%%%%%%%%%%%%%%%%%%%%%%%%%%%%%%%%%
%%%%%%%%%%%%%%%%%%%%%%%%%%%%%%%  ACKNOW %%%%%%%%%%%%%%%%%%%%%%%%%%%%%%
%%%%%%%%%%%%%%%%%%%%%%%%%%%%%%%%%%%%%%%%%%%%%%%%%%%%%%%%%%%%%%%%%%%%
\subsection*{Acknowledgments}
\noindent
The author would like to thank Yuji Hirota for useful comments and discussion.
This work was supported by the research promotion program for acquiring grants in-aid for scientific research(KAKENHI) in Ritsumeikan university.

%%%%%%%%%%%%%%%%%%%%%%%%%%%%%%%%%%%%%%%%%%%%%%%%%%%%%%%%%%%%%%%%%%%%%
%%%%%%%%%%%%%%%%%%%%%%%%%%%%%%   APPENDIX    %%%%%%%%%%%%%%%%%%%%%%%%
%%%%%%%%%%%%%%%%%%%%%%%%%%%%%%%%%%%%%%%%%%%%%%%%%%%%%%%%%%%%%%%%%%%%%
\appendix
\section{Local coordinate expression}\label{loco}
\subsection{Courant algebroid}
Let $E$ be a Courant algebroid over a smooth manifold with 
the inner product $\bracket{-}{-}$, the anchor map $\rho:E \rightarrow TM$, 
and the Dorfman bracket $[-,-]_D$.

A local coordinate on $M$ is $x^i$, and a basis of the fiber of $E$ is $e_a$.
We denote $k_{ab} = \bracket{e_a}{e_b}$, $\rho(e_a) = \rho^i_a \partial_i$,
and $[e_a, e_b]_D = k^{cd} f_{dab} e_d$.

Local coordinate expression of conditions of the Courant algebroid are 
\begin{align}
& k^{ab} \rho^i_a \rho^j_b =0,
\label{CAidentity1}
\\
& \rho^j_b \partial_j \rho^i_a - \rho^j_a \partial_j \rho^i_b
+ k^{ef} \rho^i_e f_{fab} =0,
\label{CAidentity2}
\\
& \rho^j_d \partial_j f_{abc} - \rho^j_a \partial_j f_{bcd}
+ \rho^j_b \partial_j f_{cda} - \rho^j_c \partial_j f_{dab}
+ k^{ef} f_{eab} f_{cdf} + k^{ef} f_{eac} f_{dbf} + k^{ef} f_{ead} f_{bcf} 
=0.
\label{CAidentity3}
\end{align}

%%%%%%%%%%%%%%%%%%%%%%%%%%%%%%%%%%%%%%%%%%%%%%%%%%%%%%%%%%%%%%%%%%%%
%%%%%%%%%%%%%%%%%%%%%%%%%%%%%%%%%%%%%%%%%%%%%%%%%%%%%%%%%%%%%%%%%%%%
%%%%%%%%%%%%%%%%%%%%%%%%%%%%%%%%%%%%%%%%%%%%%%%%%%%%%%%%%%%%%%%%%%%%
\subsection{Connection, curvature and torsion}\label{RTS}
Let $D:\Gamma(E) \rightarrow \Gamma(E \times T^*M)$ be a connection
and $\Gamma_a^b = \Gamma_{ai}^b dx^i$ be a connection 1-form of $D$.
The covariant derivatives with respect to $D$ is
for $\mu^a e_a \in \Gamma(E)$ and $\mu_a e^a \in \Gamma(E^*)$ are
\begin{eqnarray}
D_i \mu^a &=& \partial_i \mu^a + \Gamma_{bi}^b \mu^b,
\\
D_i \mu_a &=& \partial_i \mu_a \rc{-} \Gamma_{ai}^b \mu_b.
\end{eqnarray}
The $E$-covariant derivatives are
\begin{eqnarray}
{}^E D_a \mu^b &=& \rho^i_a (\partial_i \mu^b + \Gamma_{ci}^b \mu^c),
\\
{}^E D_a \mu_b &=& \rho^i_a (\partial_i \mu_b - \Gamma_{bi}^c \mu_c).
\end{eqnarray}
Local coordinate expressions fo the curvature, the $E$-torsion,
and the basic curvature are
\begin{eqnarray}
R_{ija}^b &=& \partial_i \Gamma_{aj}^b - \partial_j \Gamma_{ai}^b
- \Gamma_{ai}^c \Gamma_{cj}^b + \Gamma_{aj}^c \Gamma_{ci}^b,
\\
T_{abc} &=& f_{abc} - \frac{1}{2} (\rho^i_{a} \Gamma^d_{bi} k_{dc}
+ \mathrm{Cycl}(abc))
\nonumber \\
&=& f_{abc} - (\rho^i_{a} \Gamma^d_{bci} 
+ \mathrm{Cycl}(abc)),
\\
S_{jab}^c &=& D_j T_{ab}^c + \rho^i_a R_{ijb}^c - \rho^i_b R_{ijb}^a,
\end{eqnarray}
where $D_j$ is the covariant derivative on $\Gamma(E \otimes \wedge^2 E^*)$.

\newcommand{\bibit}{\sl}

%%%%%%%%%%%%%%%%%%%%%%%%%%%%%%%%%%%%%%%%%%%%%%%%%%%%%%%%%%%%%%%%%%%%%
%%%%%%%%%%%%%%%%%%%%%%%%%%%%%%%  Refs. %%%%%%%%%%%%%%%%%%%%%%%%%%%%%%
%%%%%%%%%%%%%%%%%%%%%%%%%%%%%%%%%%%%%%%%%%%%%%%%%%%%%%%%%%%%%%%%%%%%%
%\newpage
%NEW MACRO FOR BIBLIOGRAPHY

%\section*{References}
%\noindent

\end{document}